# Multiscale charge transport in van der Waals thin films: reduced graphene oxide as case study

*Alessandro Kovtun,[†] Andrea Candini,[†] Anna Vianelli,[‡] Alex Boschi,[†] Simone Dell'Elce,[§] Marco Gobbi,[⊥,¥,♠] Kyung Ho Kim,[∇,△] Samuel Lara Avila,[∇] Paolo Samorì,[⊥] Marco Affronte,[♣] Andrea Liscio,[♦]\* Vincenzo Palermo[†,♥]\**

[†] Consiglio Nazionale delle Ricerche, Istituto per la Sintesi Organi e la Fotoreattività, (CNR-ISOF), via Gobetti 101, Bologna, Italy

[‡] MISTER Smart Innovation, via Gobetti 101, 40129 Bologna, Italy

[§] Graphene-XT srl, via D'Azeglio 15, 40123 Bologna, Italy

[⊥] Université de Strasbourg, CNRS, ISIS, 8 allée Gaspard Monge, 67000 Strasbourg, France

[¥] CIC nanoGUNE BRTA, Tolosa Hiribidea 76, 20018 Donostia – San Sebastian, Spain

[♠] IKERBASQUE, Basque Foundation for Science, Plaza Euskadi 5, 48009 Bilbao, Spain

[∇] Chalmers University of Technology, Department of Microtechnology and Nanoscience, Kemivägen 9, 41296 Gothenburg, Sweden

[△] Physics Department, Royal Holloway, University of London, Egham Hill, Egham, Surrey TW20 0EX, UK

♣ Dipartimento di Scienze Fisiche, Informatiche e Matematiche (FIM), via Giuseppe Campi 213/a, 41125 Modena, Italy

♦ Consiglio Nazionale delle Ricerche, Istituto per la Microelettronica e Microsistemi, (CNR-IMM), via del Fosso del Cavaliere 100, 00133 Roma, Italy

♥ Chalmers University of Technology, Department of Industrial and Materials Science, Hörsalvägen 7, 41296 Gothenburg, Sweden

ABSTRACT: Large area van der Waals (vdW) thin films are assembled materials consisting of a network of randomly stacked nanosheets. The multi-scale structure and the two-dimensional nature of the building block mean that interfaces naturally play a crucial role in the charge transport of such thin films. While single or few stacked nanosheets (*i.e.* vdW heterostructures) have been the subject of intensive works, little is known about how charges travel through multilayered, more disordered networks. Here we report a comprehensive study of a prototypical system given by networks of randomly stacked reduced graphene oxide 2D nanosheets, whose chemical and geometrical properties can be controlled independently, permitting to explore percolated networks ranging from a single nanosheet to some billions with room temperature resistivity spanning from $10^{-5}$ to $10^{-1}$ Ω·m. We systematically observe a clear transition between two different regimes at a critical temperature T*: Efros-Shklovskii variable range hopping (ES-VRH) below T* and power law (PL) behavior above. Firstly, we demonstrate that the two regimes are strongly correlated with each other, both depending on the charge localization length ξ, calculated by ES-VRH model, which corresponds to the characteristic size of overlapping $sp^2$ domains belonging to different nanosheets. Thus, we propose a microscopic model describing the charge transport as a geometrical phase transition, given by the metal-insulator transition associated with the percolation of quasi-1D nanofillers with length $\xi$, showing that the charge transport behavior of the networks is valid for all geometries and defects of the nanosheets, ultimately suggesting a generalized description on vdW and disordered thin films.

The development of cheap techniques to produce and to process large quantities of monoatomic thick materials such as graphene[1] and related 2D-materials allowed to create synthetic structures, *i.e.* van der Waals (vdW) materials, with tuned properties.[2-3] Using solution-processing approaches,[4-6] the produced nanosheets can be arranged forming macroscopic vdW thin films composed of billions of 2D sheets randomly stacked, in which the structural and electrical continuity of the 2D/3D architectures is provided by the contact regions between the sheets. The versatility and the processability of such assembles allow to control with great flexibility both the crystal quality of the sheets and their stacking geometry (*e.g.* thin films, membranes, foams, *etc.*), holding a significant potential for the emerging applications based on large-area flexible and wearable devices,[7] coatings and advanced composites.[8]

Despite of the great technological developments, a general framework describing the charge transport (CT) in vdW thin films is still lacking. Most of the studies carried out during the last 15 years provided an in-depth insight into CT in individual almost defect-free nanosheets or stacks obtained by superimposing only a finite number of them fin the so-called vdW heterostructures.[9]

The CT in a network differs intrinsically from that of the single sheets because of the major role played by inter-sheet processes. A model taking into account the superimposition of individual sheets represents an over demanding computational task since it would require a detailed description of the mechanisms involved at interfaces on different length and time scales. Moreover, due to the need of finding the ideal trade-off between costs of production on large scale and competitive properties, the individual sheets forming the film necessarily possess intrinsically lower quality than the pristine (greater amount of defects and chemical functionalization, higher size polydispersity), thus further increasing the resources required for their simulation.

The experimental approaches used so far suffered primarily from the poor processability of graphene. In particular, macroscopic films made by exfoliated graphene dispersions contain significant amounts of multilayers. The presence of residual contaminants or dopants is also a factor that limits the ability to develop comprehensive models. Instead, the use of pure 2-dimensional, monoatomic nanosheets as (semi)conductors would enable to unravel the correlation between system dimensionality, nanostructuring of the interfaces and charge transport in complex networks.

It is worth taking into account that networks of 2D materials obtained by overlapping sheets are characterized by 2D interfaces which can be as large as the area of the individual sheets. Hence, each of the atom composing the material will lie at the interface, in contact with the sheets above or beneath it (figure 1a), being in sharp contrast with conventional 3D assemblies or composite materials. Networks of stacked 2D nanosheets can thus be defined as a multi-junction material where the CT mechanisms acting on the single sheet are strongly involved with those at the interface (sheet-to-sheet).

As prototypical material we used single monolayer sheets of reduced graphene oxide (RGO), which consists of a (semi)conductive graphene lattice comprising oxygen-containing defects.[10-11] Differently from solution-processed exfoliated graphene, such sheets can be obtained in the monodisperse form of single layers with their lateral size tunable from >100 nm to ≈ 100 μm by sonication treatments.[12] These nanosheets can thus be dispersed in various liquids, processed varying their size, film thickness hence their conductivity,[13] and easily arranged on a substrate forming networks with partially ordered structure, where all the single sheets are randomly distributed in the plane orientation and stacked perfectly parallel at a fixed distance (*i.e.* RGO interlayer distance) forming thus RGO thin films. In addition to the study of the CT mechanisms of vdW thin films, the capability to tune the chemico/physical properties of the building block and as well as to fabricate large-area networks in controlled way allowed to use RGO networks as test-bed systems to investigate the CT mechanisms in disordered and amorphous materials and as well as to verify the wide range of theories developed; see [14] and the references within for a comprehensive overview.

Here we present a scale-independent model for the charge transport in highly disordered networks of defective 2D materials based on a robust data analysis of the electrical resistivity *vs* temperature $\rho(T)$. This resistivity is a reliable physical observable to achieve in depth insight into single sheet to large-area aggregates. Moreover, we show that CT properties of such network can be described in a more general framework of the geometrical phase transition. By taking full advantage of the controlled structural and morphological composition of the films we carried out a multiscale quantitative analysis to elucidate the different transport regimes taking place in such materials. In particular, we investigated the CT mechanisms occurring at the sheet-to-sheet interfaces, typically considered as bottlenecks, as well as the role of the geometrical complexity of the network on the overall electrical conductivity of the nanosheets assemblies. The use of RGO enabled us to tune independently *i)* the conductivity of each nanosheet, *ii)* the lateral size of the nanosheets, and *iii)* the thickness of the material, *i.e.* the number of nanosheets stacked over each other.

RESULTS AND DISCUSSIONS

The starting material were water solutions of single-layer GO nanosheets. These solutions were spun onto SiO2 substrates at different concentrations, followed by a sample's reduction *via* thermal annealing at increasing temperature ($T_{ann}$ = from 200 °C up to 940 °C) for 60 minutes at a vacuum of $10^{-6}$ mbar to transform them in conductive RGO. Finally, gold electrodes were deposited by thermal evaporation through a shadow mask. Upon changing systematically such experimental parameters, a set of 28 different devices was produced. At lower nanosheet density, micrometer sized highly-reduced RGO sheets ($sp^2$ content = 96±2% as determined by X-ray Photoelectron Spectroscopy) were connected by rare and clearly visible contact points with a size to form networks on micrometer scale (figure 1b). Differently, at higher deposition density, continuous layers of stacked sheets (figure 1c), were produced with tunable oxidation degrees, sheet size tuned from <0.3 μm to >17 μm, and film thickness from 1 up to 35 layers. All the fabricated samples (a.k.a. devices) were studied by microscopic and spectroscopic techniques (see Methods) at each step of preparation.

Preliminary Field-Effect Transistor (FET) measurements performed on single and sparse network of micrometer sized highly-reduced RGO sheets (figure S5) showed hole transport with linear region mobility ($\mu_{FET,lin}$) of ca 1 – 10 cm$^2$·V$^{-1}$·s$^{-1}$ (figure 1e). Despite of the analysis of the $\mu_{FET,lin}$ vs temperature is the most common approach to study CT mechanism, such physical observable could not be the most useful to compare systems with different geometries and scale-lengths. Moreover, thin films with channel length and width of ca. 1 cm are strongly affected by the edge effects. For this reason, we focussed our attention to the study of the electrical resistivity vs temperature $\rho(T)$ for each device from ca 10 K to 300 K, to detect the existence of different transport regimes. In all the 28 different devices we found that $\rho$ decreases with the increasing of temperature, clearly indicating that all the networks possess a semiconducting behavior, with room temperature resistivity $\rho_{RT}$ lying in the range between 2 and 2·10$^{-5}$ Ω·m. All the acquired $\rho(T)$ curves are reported in the Supporting Information.

**Observation of CT regimes at different temperatures.** Typically, CT in disordered and amorphous organic semiconductors is modeled as hopping transport between localized states. Although the literature related to the transport phenomena is enormously rich and several models have been developed to study this physical framework, we can identify two "common" functional dependences that were used and combined to describe the measured ρ(T) trends: i) a stretched exponential and ii) a power law (PL). The first form is used to describe Variable Range Hopping (VRH) models[16-17] and Fluctuation-Induced Tunneling (FIT)[18] transports, typically observed at low temperatures. The latter form is observed in a wide range of systems and temperatures and it is related to high-density conductivity of disordered semiconductors (see Ref. [8] and the references within) (SI, Section 2.1).

From the experimental and data analysis point-of-view, a critical issue is related to the fitting analysis since the use of stretched exponential or PL functional forms describing ρ(T) can lead to systematic errors. For example, the application of the commonly used *least-squares procedures* for fitting a Poisson-distributed data is known to lead to biases.[19] We avoided possible artefacts by devising a robust self-consistent method based on the reduced activation energy (W), *i.e.* the logarithmic derivative of resistivity *versus* temperature:[20-21] $W(T) = -d(ln\rho)/d(lnT)$. This mathematical function transforms stretched exponentials and PL curves into linear functions in log-log space, clearly simplifying the issues related to the fitting procedures.

We should consider if deviation from linearity could be ascribed to the presence of multiple transport mechanisms acting simultaneously, whose average would (by a fortuitous chance) give a critical exponent =0.5. In such case, the resistivity ρ should be described as the sum of different terms, one for each mechanism: $\rho = \rho_1 + \rho_2 + \cdots + \rho_n$. Then, the logarithm derivative of the resistivity, W(T) will have a complicated analytic form as there is no closed-form expression to solve a sum within the logarithm argument: $W = -\frac{\partial \ln \rho}{\partial \ln T} = -\frac{\partial}{\partial \ln T} \ln(\rho_1 + \rho_2 + \cdots + \rho_n)$, thus showing a complex behaviour at least in some of the 28 devices we analysed. Conversely, all the devices showed a linear trend, thus indicating that a single CT process is in action for each given temperature.

From the best of our knowledge the W(T) method is commonly used for the study the CT of inorganic semiconductors. It has not been employed systematically for graphene-based materials; previous works used it to study conductivity only in limited, narrow portions of the measured temperature range. Currently, several VRH models have been proposed to describe the CT of graphene-based materials without a general consensus.[4, 8, 20-26] See Section S2.2 for more details.

It is noteworthy to underline that most of the analysis performed in literature are based on simple fits of ρ(T). We use the W(T) method to explore all the 28 devices proposing this systematic approach to settle this open debate.

The simple analysis of ρ(T) did not allow to compare quantitatively our samples. Differently, W(T) trends clearly revealed the differences between samples, as depicted in figure 2 by showing set of original data corresponding to a single RGO sheet (figure 2a) and a representative case of RGO network (figure 2b), either represented in the standard way ρ(T), and the corresponding W(T) trends. In both cases W(T) curves clearly showed linear trends for well-defined temperature ranges supporting the presence of single mechanisms acting at different temperatures. In the case of individual RGO sheets the log-log plot of W(T) is linear with a negative slope corresponding to a VRH regime over the entire range of the measured temperatures. Differently, for RGO networks we can clearly distinguish two regimes separated by a transition temperature ($T^*$): a linear trend with negative slope similar to that measured in single sheets for T<T* and a constant trend corresponding to PL for T>T*. Comparing all the RGO networks, we observed that T* increases with the electrical resistivity at room temperature, $\rho_{RT}$ (figure 3a, the dashed curve is a guide for eyes). A third type of regime is observed in RGO films with high reduction and high thickness (thickness > 6 nm). This further regime is present at relatively high T (> 250 K, see figure S7) and shows a positive trend of W (*i.e.* $\frac{dW}{dT} > 0$). Although a robust quantitative analysis of the experimental data is not possible due to the low signal-to-noise ratio of the ρ(T) data, the positive W trend clearly indicates that semiconductive RGO films tend to become quasi-conductive as we increase the film thickness. A similar behavior was already observed by Shaina *et al.*[27] in thick RGO films (thickness > 500 nm) with larger variations of ρ and W values showing how such regime becomes predominant when films are subjected to uniaxial strain. We could propose a single scheme summarizing the CT of RGO networks by the quantitative analysis of the two regimes comparing systematically all the 28 devices, where the single sheet is a particular case of the RGO networks where the transition temperature T* is out of the range of measured temperatures (T* > 300 K). This scheme is supported by the behavior observed on micro-networks of few partially overlapped nanosheets where T* (open circles) is compatible with the room temperature within the experimental error bar (T* ≈ 300 K) indicating that T* is related to the structural complexity of the network (*e.g.* number of overlapping nanosheets). For the sake of simplicity, the observation of linear W(T) trends in all the 28 devices clearly excludes the presence of multiple mechanisms, and allows a clear assignment of the CT mechanism. Thus, the temperature dependence of the electrical resistivity of all the measured devices – and the corresponding W(T) – could be written as:

$$\rho(T) = \begin{cases} \rho_{0,VRH} \cdot exp\left\{\frac{T_0}{T}\right\}^{\beta} \\ \rho_{0,PL} \cdot \left(\frac{T}{T_1}\right)^{-m} \end{cases} \Leftrightarrow \ln W(T) = \begin{cases} -\beta \cdot \ln T + \Delta & T < T^* \\ \ln m & T > T^* \end{cases} \quad (1)$$

where $\Delta = \ln\left(\beta \cdot T_0^{\beta}\right)$.

Both VRH and PL regimes showed similar functional forms being defined by three parameters: a prefactor ($\rho_{0,VRH}$ or $\rho_{0,PL}$), a characteristic temperature ($T_0$ or $T_1$), and a characteristic exponent ($\beta$ or $m$).

In the following paragraphs, we describe the role and the physical details of such parameters (*i.e.* the six ones defined in eqn.1 and T*) using a generalized description of the experimental features and correlating the data obtained in all the RGO networks. The data analysis procedure was performed as follows: first, we fitted the W(T) curves by calculating $\beta$, $T_0$ and $m$ as free parameters, then we set these parameters and calculated the resistivity prefactors by fitting the $\rho$(T) curves. Finally, we fitted both W(T) and $\rho$(T) using the parameters previously calculated as initial values, verifying their convergence on successive reiterations.

*VRH regime.* In VRH the stretching exponent $p$ is strongly dependent on the shape of density of states at Fermi energy, $g(E_F)$. As reported in figure 3b, all the devices showed the same $\beta$ values = 0.52±0.06, being in excellent agreement with results obtained on single RGO sheets[28-29] and with the values measured on hydrogenated graphene ($\beta$ = 0.47 – 0.58).[30]

All the $\beta$ values were always close to 1/2, providing an unambiguous answer to the long-debated nature of the charge transport in RGO thin films. For sake of comparison, figure 3b also shows the values expected for other regimes (dashed blue lines): VRH in 2D and 3D ($\beta$ = 1/3 and $\beta$ = 1/4, respectively), Nearest-Neighbours Hopping (NNH) and Fluctuation-Induced Tunneling (FIT) ($\beta$ = 1). All of such regimes can be clearly excluded. The value $\beta$ = 1/2 can be explained by two possible VRH models: 1D-Mott and Efros-Shklovskii (ES). These two models differ in the shape of the density of states: while $g(E_F)$ is constant in the former, ES is caused by a gap at the Fermi level due to the Coulombic interactions between the excited state above the $E_F$ and the hole left by the same electron below.[31] 1D-VRH is typically used to describe the CT of 1D conductors (*e.g.*, quantum wires, nanotubes or polymers),[32] but is apparently insignificant in the case of the carbon network structures.[33] Conversely, ES-VRH dominates at all measurable temperatures in case of high-disorder systems and it has been already observed in RGO sheets.[28]

The characteristic temperature of the hopping mechanism $T_0$ is inversely proportional to the localization length ($\xi$) defined as the average spatial extension of the charge carrier wavefunction. According to the ES-VRH regime, the relation is given by $T_0 = \frac{2.8\, e^2}{4\pi\, \epsilon_0 \epsilon_r k_B \cdot \xi}$,[31] where $e$ is the electron charge, $\varepsilon_r$ the dielectric constant, and $\varepsilon_0$ the electric permittivity of the vacuum. For sake of simplicity, we compacted the equation as: $T_0 = \delta/\xi$, where $\delta$ collected all the constant values. Concerning the dielectric constant, we should consider that charges will hop form one nanosheet to the other during the mesoscopic transport, changing continuously their depth inside the RGO layer. Thus, all charges will be exposed part of the time to the outer vacuum, part of the time to the underlying SiO2, and always to RGO. The effective dielectric constant felt by the charge will depend on how much time the charge will spend in the outer or inner layers; however, except from the extreme cases of a perfect monolayer compared to macroscopic bulk RGO, the average dielectric constant felt by the charge in few-layers RGO should be comparable. In the case of single of few sheets we estimated $\varepsilon_r$ = 2.5, resulting from the average of the dielectric constant of the substrate (SiO2 = 3.9) and of the medium (vacuum/air = 1),[34] while in the case of RGO thin film we considered $\varepsilon_r$ = 3.5, according to previous reports.[35-36] ES-VRH regime fully describes the measured ρ(T) behavior of single RGO, PL regime is instead dominant at high temperature (close to RT) when two or more nanosheets are assembled partially in contact with each other. The PL regime describes ρ(T) in most of the measured temperature range for thick RGO films. Thus, the observation of the PL regime indicates a clear fundamental role of the inter-sheet CT in a network, as detailed below.

Similar results were observed in different GRM thin films. Recently Silverstein *et al.*[37] exploited the W(T) approach to investigate CT of RGO thin films obtained by voltage-induced reduction clearly observing two regimes separated by a transition temperature T*: a VRH regime with $\beta$ = 0.43 for T<T* and a roughly constant trend (slope = 0.03) for T>T* suggesting a PL regime. Taking into account the confidence intervals calculated by digitizing the published data both values show an excellent agreement with the scheme here presented. A further agreement with the two-regimes scheme is reported by Turchanin *et al.*[38] in the case of GRM thin films obtained by cross-linking of aromatic molecules. Although no W(T) analysis were available for such data, the resistivity showed a VRH behavior for few layers with a transition to PL with the increasing thickness. This analogy with systems that are less disordered than the one discussed here suggests that the interplay of the two regimes may describe the microscopic properties of assembly of 2D nanosheets in general, and not only RGO materials.

*PL regime.* We performed further analysis to elucidate the CT mechanism occurring in the PL regime. Furthermore, we wanted to understand if the ES-VRH and PL regimes are correlated and if it is possible to use only a single parameter to describe the electrical properties of all the RGO networks, for the whole temperature range studied.

Differently from the VRH regime, in PL regime only m values could be evaluated using the W(T) analysis (figure S7-S14). The m values observed ranged between 0.2 and 4.0 (table S3) increasing with the measured $\rho_{RT}$ of each network (Figure 3c) and the PL characteristic temperature $T_1$ could not be directly decoupled from the prefactor $\rho_{0,PL}$. The resistivity measured in the PL regime is often described, in literature using a general function $\rho = B \cdot T^{-m}$, where $B$ is the scale factor. We could combine such general function with eqn. 1 to obtain:

$$\ln B = \ln \rho_{0,PL} + m \cdot \ln T_1 \qquad (2)$$

It is noteworthy to underline that eqn.2 is a mathematical equivalence between the parameters of each individual *i-th* RGO network. Therefore, we should consider a sequence of independent equivalences, one for each measured sample, where *m* and $\ln B$ parameters are directly calculated by W(T) and ρ(T) analysis, respectively. Moreover, each equivalence showed a solution given by a linear combination of $\ln T_1$ and $\ln \rho_{0,PL}$.

Figure 3d shows the correlation plot $\ln B$ *vs m* for each sample. If the parameters were uncorrelated, randomly scattered data should be observed in the plot. Conversely, all data lay on a straight line (red line), as depicted in figure 3d. Such experimental evidence allowed i) to decouple the characteristic temperature $T_1$ and the prefactor $\rho_{0,PL}$ and ii) to prove that such parameters are the same for all the samples, corresponding to the constant slope ($\ln T_1$) and the Y-intercept ($\ln \rho_{0,PL}$), as calculated by a linear fit according to eqn.2. Considering the PL functional form reported in eqn.1, we achieved for all the RGO networks $k_B T_1 = 148 \pm 5$ meV, $\rho_{0,PL} = (4.4 \pm 0.1) \cdot 10^{-6}$ Ω·m. The invariance of such parameters clearly indicates that the same mechanism governed the CT of all the RGO networks. The power exponent *m* is the only parameter that independently identifies each sample. We compared the measured *m* values with those reported in literature for conjugated polymers to cast light on the mechanism involved in the CT. The functional PL shape ruling the relation between electrical resistivity and temperature corresponds to the *critical regime* of the Metal-Insulator Transition (MIT) in conducting polymers.[39-40] The MIT model described the conductivity behavior being neither metallic nor insulating, predicting a value of *m* between 1/3 and 1.

Another approach is based on the use of a *universal scaling curve*,[41] being such curve commonly observed in low-defects semiconductive polymers. The CT along the polymeric backbone is influenced by the coupling strength of the charge and the nuclear vibrations (*i.e.* a phonon bath). This mechanism is described by the quantum nuclear tunneling (NT) theory, in which the carriers tunnel through the potential barrier formed by the coupling of the electronic charge to its nuclear environment. Such coupling constant is called Kondo parameter, defined as $\alpha_K = \frac{m}{2} + 1$.[41] The higher the value, the stronger the coupling. Typical values reported in literature in π-conjugated polymers for bath phonon energies are ~100 meV[42] and $\alpha_K$ within the range 1.6 – 6.75,[43] corresponding to *m* values between ca. 1.2 and 11.5. An alternative description of PL transport, assuming multiphonon tunnelling of localized electrons with a weak electron-lattice interaction,[44] could be discarded because in our case the phonon bath energy was >> $k_BT$. We found that the power exponent *m* in RGO networks spans between the ranges defined by the values calculated by MIT and measured using NT models (Figure 3c). Such experimental evidence i) highlight the analogy between PL transport in RGO networks and conjugated polymers thin films, and ii) suggest that both classes of materials can be studied with similar approaches. Thus, we could describe the parameter *m* similarly to the case of polymers, in terms of reorganization energy (λ) of phonon bath upon electron transfer.[45] We generalize the NT theory defining the phonon bath energy in RGO networks = $k_B T_1$. Thus, the reorganization energy term describes the strength of the electron-phonon bath and can be reliably estimated as twice the relaxation energy of a polaron localized over the region of two overlapped RGO sheets, *i.e.* the twice the polaron binding energy: $\lambda = \left(\frac{m}{2} + 1\right) \cdot k_B T_1$.[46]

**Correlation between ES-VRH and PL regimes.** A natural question to be addressed is whether the two CT regimes that we observed at low and high T (ES-VRH and PL) are correlated. We observed that:

i) the PL exponent *m* (or the correlated reorganization energy λ) decreases with the increase of the localization length calculated in ES-VRH (figure S10), and

ii) the measured ρ(T) data and the corresponding W(T) analysis do not show discontinuities at the transition temperature T* between ES-VRH and PL regimes.

The first experimental evidence confirms a strong analogy with polymeric semiconductors, where the reorganization energy decreases as the size of the π conjugated system is increased.[47] Moreover, the measured trend (Figure S10) roughly agrees with the relation $\xi \propto 1/m^2$ computationally obtained for highly doped 1D semiconductors by Rodin *et al.*[48] The second experimental result implies that, at the transition temperature T*, the resistivity calculated in the VRH regime corresponds to that in PL one: $\rho_{VRH}(T^*) = \rho_{PL}(T^*)$. As shown in the study of the functional forms of the ρ(T) curves, also in this case we used the W(T) method focusing the analysis of the equivalence achieved for W curves: $W_{VRH}(T^*) = W_{PL}(T^*)$. The use of linearized curves made it possible to derive a general formula comparing the parameters of the two CT regimes: *T**, *β*, *m* and *ξ*, as calculated using eqn. 1 for each *i-th* RGO network. Calculating all the terms as sum of logarithms (for more details see SI, Section S6.1), we obtained:

$$\beta \cdot \ln T^* + \ln m = -\beta \cdot \ln \xi + \beta \cdot \ln \delta + \ln \beta \qquad (3)$$

Similarly to the correlation analysis discussed before in eqn. 2 and figure 3d, we should consider a set of independent equations, one for each measured *i-th* RGO network. Figure 3e shows the correlation plot $\ln \xi$ *vs* $(\beta \cdot \ln T^* + \ln m)$ where each point corresponds to the data of each RGO network and the red line corresponds to a linear fit from eqn. 3. The plot shows an excellent linear behavior, with slope = $-\beta_{mean}$ = -0.55±0.03, in good agreement with *β* = 1/2 (ES-VRH). The good correlation achieved is an additional proof that ES-VRH and PL regimes should be strongly correlated, even if the two regimes are based on different mechanisms and act at different temperature ranges. A similar approach related to equivalence of the resistivity values is reported in SI (figure S12 and Section S6.2).

A clarification is needed on the physical meaning of the transition temperature T*. Experimentally, T* values range between 200 K and 20 K with the increase of the localization length (Figure S11). This is in qualitative agreement with the model developed for disordered wires by Gornyi *et al.*,[49] where T* is related to the disorder and it is inversely proportional to the localization length. Noteworthy, the model previously proposed by Rodin,[48] where the PL exponent *m* is proportional to the number of hopping events, can be seen as a special case of eqn. 3, assuming T* as constant and $\beta_{mean} = 1/2$.

**Dependence of ξ on the structural and geometrical properties of the RGO nanosheets network.** In the case of single RGO sheet the transport is purely 2D, the localization length ξ corresponds approximately to the half of the size of the aromatic domains ($\phi$): $\phi \approx 2\xi$,[28] and charges "hop" from one domain to the next one. Differently, when two or more sheets are overlapped (even partially) we can distinguish two components in the transport: in-plane and out-of-plane. The case in which two adjacent sheets (*i.e.* belonging to the same plane) only touch the edges is statistically negligible. Thus, the sheet-to-sheet transport is mainly out-of-plane and consequently the in-plane component is related to the single sheet.

Experimentally, we varied such two components tuning independently different parameters which define the material: the oxidation degree (*i.e.* the corresponding intrinsic conductivity), the lateral size of the nanosheets composing the network, and the thickness of the thin film (*i.e.* the number of stacked nanosheets layers $N_{layer}$). In general, we observed that ξ increases with the decrease of the room temperature resistivity (Figure 4a). First, we varied the intrinsic conductivity of each nanosheet composing the network. This could be achieved by changing the content of the $sp^2$ carbons on the RGO basal plane by thermal annealing, as evidenced by XPS measurements (Figure 4b). The average film thickness was kept constant with the average number of stacked sheets being $N_{layer}$ = 8±1. The average lateral size of the single RGO sheets was $<s_{RGO}>$ = 428±14 nm. Second, we varied the lateral size of the RGO sheets (Figure 3b). We tested nanosheets having 3 different lateral sizes: S1 = 17.2±0.6 µm, S2= 428±14nm and S3= 380±7nm. For each size, three different $sp^2$ contents: 77% (■), 86% (■) and 96% (■) were tested for a total of 9 different size-conductivity combinations. The localization length was found to increase significantly with the aromatic content and to a lesser extent, with the nanosheet lateral size.

The results of the electrical characterizations of these films are provided in figure 4b: for low $sp^2$ content (≲80%), we obtain 1.3 < $\xi$ < 5 nm, in agreement with the values typically observed in a single RGO sheet.[11, 50] For higher aromatic contents (≳80%) the parameter $\xi$ reaches a size up to 300±20 nm evidencing that charges can be delocalized over the single aromatic region on the RGO sheet. We want to underline that such result is peculiar to the macroscopic network because in the case of an assembly of a few sheets partially in contact, the measured $\xi$ amounts to a few nm (see table S3 and figure S8, devices #26 and #27), despite the $sp^2$ content amounts 96%. Such results can be accounted for only assuming that charges travelling along the nanosheet (in-plane) prefer to circumvent a defect, jumping on different planes (out-of-plane) (Figure 5a). This suggests that the electronic states can span through different overlapping $sp^2$ domains belonging to different sheets; charges are delocalized over different nanosheets and contribute thus to CT as a single conductive domain with characteristic size $\xi$. Thus, we investigated the influence of out-of-plane mechanisms by preparing devices with an increasing number of stacked layers, *i.e.* having increasing thickness, from single RGO sheets, covering only partially the substrate, to macroscopic stacks having a high number of stacked nanosheets up to $N_{layer}$ = 34±2 (figure 4d). $\xi$ increases sharply at $N_{layer} \approx 5$, then reaching a constant value ranging between 2-3 µm for $N_{layer} \approx 8$.

For each sample, we measured the dependence of room temperature resistivity ($\rho_{RT}$) on $N_{layer}$ (figure S13) showing, as expected, an opposite trend with respect to that observed on $\xi$ in the case of RGO thin films. Such behavior is similar to that observed in thin metal films (<100 nm) where the electrical resistivity become larger as the film thickness decreases in size. In such systems this change occurs because the mean free path of charge carriers is reduced due to increased scattering effects.[51] The inter-sheet CT (out of plane-plane) is favored respect to intra-sheet CT (in-plane). This result agrees with the observation of coherent commensurate electronic states at the interface between $sp^2$ regions, recently observed.[52] In general, all the layered graphitic materials have similar electrical properties along the out-of-plane direction; for example turbostratic and Kish, high oriented pyrolytic and natural graphites usually show metallic- or semi-metallic-like behavior.[53-56]

**RGO networks as "composite materials".** The CT of RGO networks is governed by π-conjugated regions given by the overlapping $sp^2$ domains connected by a network of random paths with $\xi$ as a characteristic length. Thus, the longer the π-conjugated domains due to increased amount of aromatic content or the number of RGO layers, the greater the localization length, *i.e.* the lower are both the parameter m and the corresponding reorganization energy, as observed in semiconducting π-conjugated systems.[57]

We developed a purely geometrical approach to describe $\xi$ as a characteristic size and to reproduce as well the behavior of such parameter with the number of layers. RGO networks are described as composite materials (Figure 5b), where π-conjugated regions (*i.e.* both overlapping $sp^2$ domains in red and isolated ones in yellow) represent conductive fillers activated by an external electrical field **E**, while holes, defects and $sp^3$ insulating regions behave like an insulating matrix (green). For example, we reported the case of three-layer RGO film with randomly distributed $sp^2$ domains (disks) on each layer. We considered only transport through the stack (out-of-plane) and negligible contribution of in-plane transport to resistivity. Fillers obtained by the overlapping stacked disks are visualized by collapsing all the layers on the projected plane (red regions in bottom surface).

A similar cartoon representing the lateral view of the multilayer RGO is depicted in Figure 5c (*i.e.* disks become segments) allowing a direct visualization of the complex morphology of a representative filler (red line) given by a random path with blobs and dead ends spanning in 3 dimensions with an overall length equal to the localization length $\xi$. Using a classical physics approach, the random paths network corresponds to the superimposition of all the possible trajectories of the charge carrier that travels between the planes "jumping" from one aromatic domain to another. Geometrically, this is analogue to describe the problem of a liquid flowing in a porous structure where holes correspond to the single $sp^2$ domain. Such connectivity of holes is a tortuous conduction path which can have multiple passes and dead ends. In general, the greater the overlap of the holes (*i.e.* porosity increasing), the better the passage of the liquid.

Since the localization length becomes orders of magnitude larger than the average size of a $sp^2$ region in a single sheet: $\xi \gg \phi$, we can neglect the $sp^2$ domain size and consider only the distance between two overlapping aromatic regions (*d*) corresponding to the layer-layer distance.

Thus, we can write a general form to describe the number-of-layer -dependence of the localization length as:

$$\xi(N_{layer}) \sim \frac{\phi}{2} + d \cdot \sum_{i=2}^{N_{layer}} n(i) \qquad (4)$$

where the first term corresponds to the contribution of the single layer ($i = 1$) and the latter to the following layers. $n(i)$ is the number of steps (*i.e.* the number of conductive sp$^2$ regions involved in the path, being the stack) along the *i*-th layer, such parameter follows roughly a Poisson distribution because of the external bias is parallel to the film ($\boldsymbol{E} = E_\parallel$).

Given a random path in (3D) thick film, the decrease of layers corresponds to a reduction of the number of steps. When the film thickness decreases down to a critical thickness (in our case, $N_{layer} \approx 8$) the continuous percolated random path breaks in small connected regions, corresponding to an abrupt decrease of $\xi$. The ultimate case corresponds to a single sheet ($N_{layer} = 1$), where no sp$^2$ regions are overlapped and $\xi$ roughly corresponds to two times the typical sp$^2$ domain size in RGO (<10 nm, table S3). The fitting curve obtained by eqn.S6.3 (red line in Figure 4d, see SI Section S6.3) shows an excellent agreement with the experimental data.

**Disorder and Phase Transition.** Dispersed in the form of a highly entangled (interconnected) structure in the matrix, the conductive fillers exhibit distinct curved and branched shapes with total length equal to $\xi$, ultimately forming an interlocked structure of quasi 1D-fillers in the agglomerated state (a.k.a. *spaghetti-like structure*). The longer the conductive fillers, the better is the electrical conducibility of the system, in agreement with the experimental findings where $\xi$ increased with the decreasing of the measured room temperature resistivity ($\rho_{RT}$) (Figure 6a). The results described above indicate that the same CT mechanisms governed all the devices tested and did not depend on the size of the nanosheet network as well as the number of the RGO sheets involved in the CT (ranging from a single sheet to some billions). The absence of a characteristic length ($\xi$ varies between 1 nm to ca 3 μm) indicates some sort of "universal" behaviour of CT in terms of a percolation process at the disorder-induced metal-insulator transitions.[58] From the considerations above, it seems evident that there are clear analogies between RGO networks and conjugated polymer thin films; suggesting us to use the approach developed by Heeger to study conductive polymeric thin films.[40, 59-60] A useful qualitative indicator of the extent of disorder used in such systems is the ratio of the electrical resistivity measured at the lowest and highest measurement temperatures tested. We adopted the same approach and defined a parameter $\rho_r = \rho_{50K}/\rho_{RT}$, 50 K being the lowest temperature where all devices showed a measurable resistivity. The relation between $\rho_r$ and $\xi$ for all our devices is shown in figure 6a. As expected, the disorder decreases when the localization length $\xi$ tends to grow asymptotically. Moreover, the inset in figure 6a shows the linear correlation between the left term in eqn. 3: $(\beta \cdot \ln T^* + \ln m)$ and $1/\rho_r$, indicating that $\rho_r \propto 1/\ln \xi$, in agreement with what previously suggested by Gornyi *et al.* for low-dimensional disordered systems [49] (SI, Section 6.4).

Heeger defined $\rho_r$ as an "effective order parameter" for metal-insulator transitions (MIT) in the case of semiconducting π-conjugated polymers.[61] However, such parameter can give several issues. It is clear that this approach could not be quantitative since the arbitrary choice of the chosen temperatures. Moreover, in the case of RGO networks, $\rho_r$ roughly indicates the ratio between resistivity corresponding to VRH and PL regimes; for each RGO network the two regimes were sampled in different ways, since for each sample T* lies in different positions within the 50 K-room temperature range (see Figure 3a). For this reason, we propose to use the prefactors $\rho_{0,VRH}$ and $\rho_{0,PL}$ as the characteristic resistivity values of the two regimes and to define a "general effective order parameter" as $\rho_{0,r} = \rho_{0,VRH}/\rho_{0,PL}$. Experimentally, we achieved that i) $\rho_{0,VRH}$ tends to $\rho_{0,PL}$ with the increase of $\xi$ (Figure S14), corresponding to $\rho_{0,r} \to 1$, and ii) $\rho_{0,PL}$ is the same for all the RGO networks (Figure 3d).

We could thus describe the CT of RGO networks in terms of geometrical phase transition (*i.e.* percolation)[8] of random networks of conductive wires with mean length equal to $\xi$. The complete disorder-driven transition to metal corresponds to the case in which the critical regime is achieved for all the temperatures and the electrical resistivity no longer depends on temperature. In such case, no VRH regime could be measured and the electrical resistivity ρ(T) trend could be described by a power law (*i.e.* T* → 0K) with the exponent $m \to 0$. In the case of m = 0, the solution of the continuity equation: $\rho_{VRH}(T^*) = \rho_{PL}(T^*)$ is given by $\rho_{0,VRH} = \rho_{0,PL}$ and corresponding to $\rho_{0,r}$ = 1. The analytical solution is provided in Section S6.2. Figure 6b shows a schematic diagram of resistivity *vs* temperature in in the vicinity of MIT. Displayed as a double logarithm graph, the VRH regime is described by a non-linear curve that decreases until it becomes (at T*) a straight line (PL regime) with slope *m*. with tends to zero at the complete transition.

We modelled the disorder-driven MIT in terms of the critical exponent formalism in continuous phase transition theory, assuming that:

i) $\xi$ corresponds to the mean cluster size and is correlated to the order parameter;

ii) $\rho_{0,PL}$ corresponds to the critical threshold of the transition.

Mathematically, a percolating system is described by a parameter *p* which controls the occupancy of bonds (or sites) in the system. At a critical value $p_c$, the mean cluster size goes to infinity and the percolation transition takes place. As one approaches $p_c$, the mean cluster size diverges following a power law as $|p - p_c|^{-\gamma}$, where γ is the critical exponent depending on the dimensionality. γ= 43/18 for 2D- and γ=1.8 for 3D-systems, as reported in Table 2 in ref.[58]. Such mathematical approach is general, the parameter *p* is a probability and shows the same role as T in thermal phase transitions - a short list of examples is reported in ref.[62], Table 1.1 - or the filler loading in a composite in MIT, for instance.

We applied the percolation formula to follow the transition between the –: $\rho_{0,VRH}$ and $\rho_{0,PL}$. In our approach, *p* corresponds to $\rho_{0,VRH}$ and the percolation threshold $p_c$ is $\rho_{0,PL}$, defining the variable Π as:

$$\Pi = |\rho_{0,VRH} - \rho_{0,PL}|/\rho_{0,PL} = |\rho_{0,r} - 1| \qquad (5)$$

where at the percolation threshold $\rho_{0,VRH} = \rho_{0,PL}$ ($\rho_{0,r} = 1$), the difference is zero (Π = 0) and the mean wire length is infinite (ξ → ∞). Conversely, $\rho_{0,VRH} \gg \rho_{0,PL}$ ($\rho_{0,r} \to \infty$) and Π → ∞ when ξ → 0. The approach to the threshold is governed by power laws, which hold asymptotically close to critical value $\rho_{0,PL}$: $\xi \sim \Pi^{-\gamma}$, as a continuous transition. This expression is analogue to the magnetic fluctuations / susceptibility $\chi \sim |T - T_c|^{-\gamma}$, where $T_c$ is the threshold temperature. In the case of random networks of conductive wires (*i.e. spaghetti-like structure*) the cluster size can be easily approximated to the wire length and thus, γ is the critical exponent describing the mean wire length. Figure 6c depicts the relation between ξ and Π in log-log scale over two orders of magnitude. The observed negative linear trend confirms the power law behavior with γ = 1.7±0.2. Such result agrees with the critical exponent of 3D systems and allows to describe RGO networks as analogous to entangled disordered systems of conductive 1D fillers with fractal dimension amounting to 2.53, close to the transition point (see Table 2 in ref.[62]).

It is noteworthy to underline that the percolation approach model does not depend on nanoscopic details, such as the chemical or electronic properties of the thin-film or the physical mechanisms of CT between RGO sheets. Such approach reveals the analogy between the networks of randomly stacked 2D sheets with the networks of 1D conductive channels with complex morphology and a well-defined fractal dimension. Moreover, fundamental works of Epstein *et al.*[63] and Fogler *et al.*[32] pointed out the role the inter-chain coupling showing that the CT of networks of 1D systems is described by ES-VRH instead of 1D-VRH, as discussed above.

A similar approach could be used to model the further regime observed at high temperature with the increase of the RGO film thickness and identified by W(T) with positive trend (Figure S7, *e.g.* device #10). Observed by Govor *et al.*[33] in self-organized carbon networks and ascribed to the transfer of the charge carriers excited at the mobility edge, such transition could be described as a MIT induced by temperature, modelling the behavior of resistivity in thick RGO films subjected to uniaxial strain.[27]

CONCLUSIONS

In summary, by taking advantage of the exceptional processability and tunability of sizes and properties, RGO was used as prototypical building block to study CT in vdW thin films given by percolated networks of pure 2D nanosheets stacked over each other, yielding a (semi)conductive material with a strongly anisotropy and tunable structure. Exploring a set 28 different systems – from a single RGO nanosheet to networks of few nanosheets partially overlapped and macroscopic thin films – we could measure systematically the electrical resistivity *vs* temperature $\rho(T)$ obtaining a different behavior between networks of sheets (thin films and few-sheets partially in contact) and the single RGO sheet. While the latter only featured the Efros-Shklovskii VRH regime, the former show the presence of an additional PL regime at high temperature. The clear transition between two different yet correlated regimes makes it possible to define some key points in the long-debated CT mechanism in films obtained from solution processing of 2D materials. The charge transport and as well as the electrical behavior of the networks do not depend much on the size of the nanosheets but primarily on the sp2 aromatic clusters belonging the RGO nanosheets and how much they overlap when the nanosheets stack. Sheet-to-sheet interfaces do not act as bottlenecks; instead, the electrical behavior of the network is dominated by the inter-sheet transport through the overlapped aromatic regions describing a quasi-1D path where $\xi$ is the characteristic length. Thus, we could model the charge transport of RGO networks in terms of composite materials where quasi-1D building blocks behave as conductive fillers in an insulating matrix taking advantage of models commonly used to study charge transport in conjugated polymeric thin films and generalizing the approach developed by Alan Jay Heeger for conductive polymers in the vicinity of disorder-driven metal-insulator transition (MIT). Since $\xi$ varies over three orders of magnitude (from very few nanometer to ca 3 micron), we considered scale invariance approach describing the charge transport as a geometrical phase transition, given by the metal-insulator transition associated with the percolation of quasi-1D nanofillers.

Introducing the fractal geometry, Mandelbrot had wondered what the dimension of a ball of yarn was, and how it depended on scaling regimes.[64] This is similar geometrical issue dealt with the networks of 2D RGO nanosheets whose CT can be described in terms of bulk (3D) composite with quasi-1D conductive fillers. Based on a systematic and quantitative analysis of experimental data, the developed model combines concepts, approaches and results of different fields, such as graphene and 2D related materials, polymers science, networks, percolation and critical phenomena. The role of interfaces and their nanostructure, the universal behavior observed from the nano- to the macro- scale and the analogies with charge transport in 1D and 3D systems give to this result a broad importance, and could be used in principle to describe the CT for vdW thin films and many others disordered materials, such as composites and granular materials.

EXPERIMENTAL SECTION

**Device preparation.** Graphene Oxide (GO) was produced by modified Hummers methods, *via* the intercalation of strong oxidizing agents into Graphite, sonication in to ultrasonic bath and purified by dialysis.[65] The average size of the GO flakes was tuned by using ultrasonic bath of the solution for different amounts of time, following a previously reported procedure.[12] Thin films were produced by spin-coating the GO water suspension on a clean $SiO_2$/Si substrate (2,000 rpm for 60s), the concentration of GO in water was 2.0 g/L for the high nanosheet density and 0.3 g/L for the low nanosheet density. To tune the thickness, the procedure could be repeated multiple times (see below). Reduced Graphene Oxide (RGO) thin films were produced by thermal annealing in high vacuum ($10^{-7}$ mbar) at different temperatures the GO films. In order to prevent the potential occurrence of a short circuit between the RGO film and the doped silicon, the portion of film close to the edges was mechanically removed for a width of about 1 mm (Figure S3a). All the 28 devices showed very low leakage currents $i_{leak}$ ($10^3$ times lower than source-drain current).

28 devices were produced by changing systematically three parameters: i) $sp^2$ content, *via* thermal reduction of GO, with a $sp^2$ content ranging from 74% to 96%), ii) nanosheets with different lateral sizes from <0.3 µm to >17 µm, and iii) the film thickness, spanning from 1 up to 35 layer. We labelled each device with a progressive number, from #1 to #28: the effect of the thickness is investigated in devices #1 – #5 employing the 2 probes analysis and in devices #6 – #10 with the 4 probes geometry; the chemical composition (carbon $sp^2$ fraction) is varied in devices #11 – #16, where we re-analyzed previously published (by our group) results;[24] the flakes lateral size is varied in devices #17 – #25 together with the $sp^2$ fraction; device #26 and #27 are a double sheet RGO partially superponed and device #28 is a single sheet device. The details about each device, as size, thickness, resistivity, annealing temperature and $sp^2$ fraction are reported in SI.

Optical images of RGO few-sheet networks were acquired with Olympus BX 51 TRF.

**X-ray Photoelectron Spectroscopy (XPS).** The C 1s high-resolution XPS spectra were recorded with a Phoibos 100 hemispherical energy analyzer (Specs) using Mg Kα radiation ($\hbar\omega$ = 1253.6 eV; X-Ray power = 125 W) in constant analyzer energy (CAE) mode, with pass energies of 10 eV. Fitting the spectra in the region of the C 1s peak allowed to quantify the relative amounts of aromatic carbon (C-C $sp^2$, 284.4 eV), aliphatic carbon (C-C $sp^3$, 285.0 eV), hydroxyl (C-OH, 285.7 eV), epoxy (C-O-C, 286.7 eV), carbonyl (C=O, 288.0 eV) and carboxyl (O-C=O, 289.1 eV), (Figure S1). The XPS spectra corresponding to different reduction degrees, and the details of the fitting procedure are given in our previous publications.[66] In particular, the shape of aromatic carbon peak is chosen as asymmetric following our recent work,[67] while the other peaks are modelled by a pseudo Voigt curve. XPS analysis were used to monitor the thermal reduction of the pristine GO and as well as the reproducibility of the fabricated devices. At fixed annealing temperature, XPS spectra were acquired on RGO films with different thicknesses showing no variation in the chemical analysis confirming the uniformity of the thermal-reduction across multi-layered stacked films.

**Atomic Force Microscopy (AFM).** AFM was employed to measure the film thickness using a Bruker MultiMode 8, with probe cantilevers model RTESPA-300 (material: 0.01 - 0.025 Ωcm Sb (n) -doped Si, $f_0$: 300 kHz, k: 40 N/m, from Bruker) working in the tapping mode. Image processing was performed using the histogram analysis method.[68] Since the thickness of single layers of GO and RGO changes from 1.0 to 0.4 nm as a function of the annealing temperature,[69] we report the film thickness as the number of effective layers $N_{eff}$. (details are reported in SI, figure S2)

**Electrical resistivity measurements ρ(T).** After thermal annealing, electrical contacts were deposited on the RGO devices by thermal evaporation of 3 nm of Cr and 100 nm of Au. The device was then mounted on the device holder and connected with a micro-bonder and silver paste. In the 2 probes configuration, the channel length and channel width of the devices were estimated by using the optical microscope and are reported in table S4. In the 4 probes geometry the contacts (1 mm diameter) were evaporated in the Van Der Pauw geometry at the corners of a ≈1 cm$^2$ device. The distance between the contacts was 9 mm. Representative pictures of the devices in the two configurations are reported in figure S3b. The electrical resistance (R) was measured using the Ammeter-Voltmeter Method (Figure S4a) and the corresponding resistivity was calculated as: $\rho = R \cdot w \cdot h/L$, where w, h and L are the channel width, length and height, respectively. All the geometrical sizes are reported in Table S4. Contact resistance was measured using the Transfer Line Method (TLM) and can be neglected (Figure S4b).

Devices #26 – #28, micrometric electrodes were lithographically patterned to characterize the electrical transport across a limited number of overlapping flakes. Lithography was carried out by exposing a standard photoresist (AZ1505, Microchemicals) with the 405 nm laser of a Microtech laser writer. A 30-nm-thick Au film (without adhesion layer) was thermally evaporated onto the patterned photoresist and lift-off was carried out in warm acetone (40 °C). Relatively large flakes were chosen for these devices, leading to a channel length and width in the range 20 - 100 µm.

The resistance *vs* temperature measurements were carried out with a Quantum Design Physical Properties Measurements System (PPMS), using an external Keithley 2636 Source-Meter. The resistance was measured in the temperature range between 300 K to 2 K with a slow ramp (1 K/min). The Ohmic behavior of the device was checked by the linearity of the I-V curves (see figure S4a). In the 2-probes configuration, we fixed the voltage bias at a value in the range 25-250 mV (depending on the device resistance at room temperature) and measured the current. In the 4-probes configuration, we fixed the current in the range of 10 µA and monitored the voltage. No significant differences were observed between 2- and 4- probes measurements in the whole temperature range.

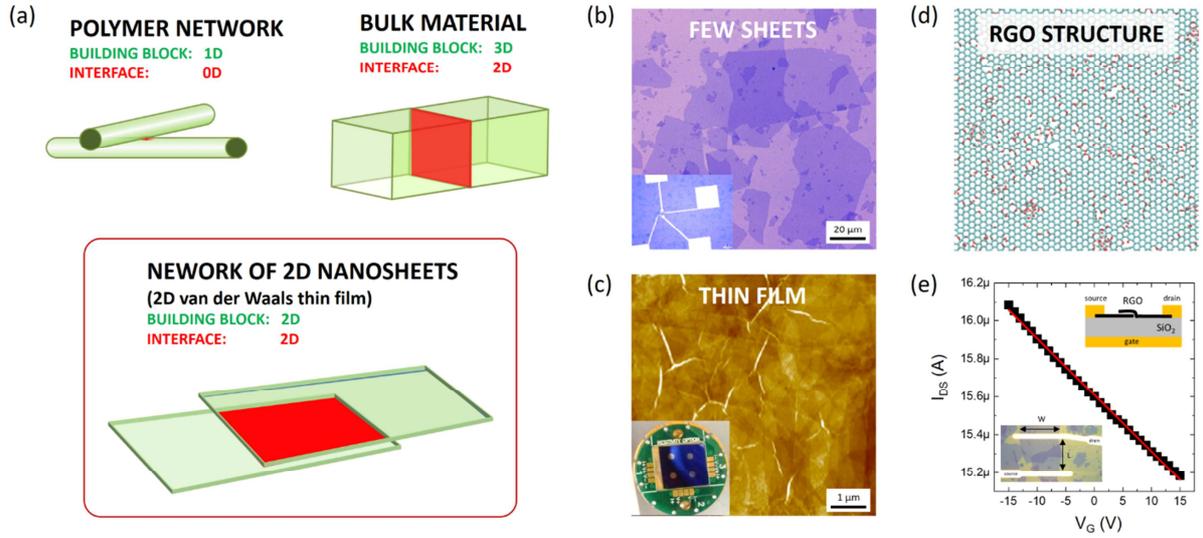

**Figure 1.** RGO networks. (a) Cartoon showing typical interfaces in systems with different dimensionality. (b) Optical image of a sparse network composed by few 2D RGO sheets in partial contact. Inset: optical image showing the metal pad geometry used to measure CT in such samples; inset size 1.7×1.2 mm$^2$. (c) Topographical AFM image of a thick RGO film (Z-range = 4 nm). Inset: photograph of the device used for the ρ(T) measurements performed at temperatures ranging from 300 K down to 5 K. (d) Structure of RGO sheet, modelled and kindly provided by Aleandro Antidormi (see[15] for details). (e) Few RGO-sheets Field-Effect Transistor. Measured transfer characteristic $I_{DS}$-$V_G$ (black symbols) and linear fit (red line) with scheme and zoom of the optical image as insets. FET mobility in linear region: $\mu_{FET,lin} = m_{lin} \frac{L}{W} \frac{1}{V_{DS}} \frac{1}{C_{ox}} \approx 2$ cm$^2$/V/s, where W ≈ L ≈ 25 μm, $V_{DS}$ = 0.5 V, $C_{ox}$ = 1.5·10$^{-8}$ F/cm$^2$ is the gate insulator capacitance per unit area and $1/m_{lin}$ = 33.5±0.4 MΩ is the inverse of the slope of the linear fit (red line).

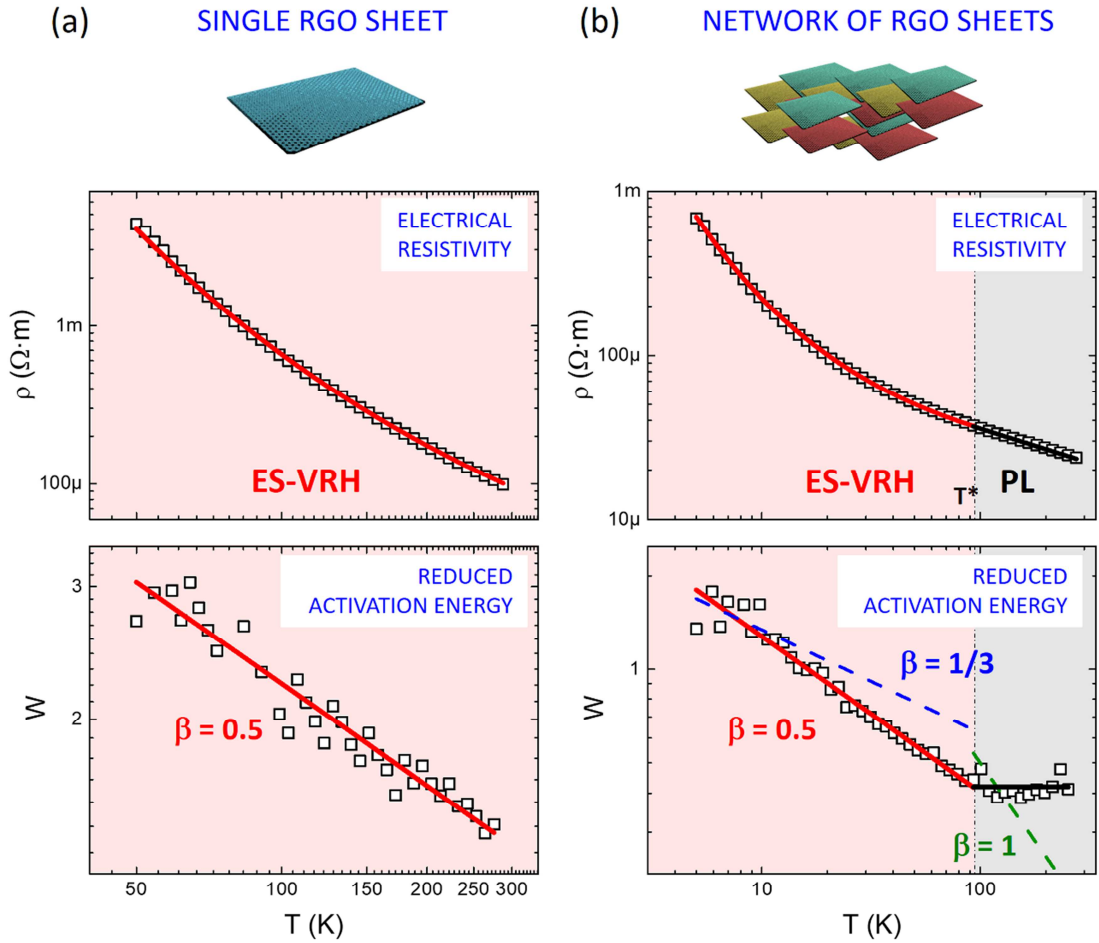

**Figure 2.** Log-log-scale plots of ρ *vs* T and corresponding Reduced Activation Energy W(T) measured in (a) single RGO sheet and (b) a typical RGO network. Two transport regimes are observed: ES-VRH (red curve) and PL (black curve). While different models can fit the measured ρ(T) trends, the use of W(T) unambiguously indicates that the correct models are ES-VRH with exponent $\beta = 0.5$ (red line, slope of W(T) linear region) at low temperature and PL (black line) at high temperatures. 2D-Mott VRH ($\beta = 1/3$) and NNH ($\beta = 1$) models are reported as blue and green dashed lines, respectively. In general, the overall W(T) trend measured on all the devices is a concave function (see SI, paragraph 3.1).

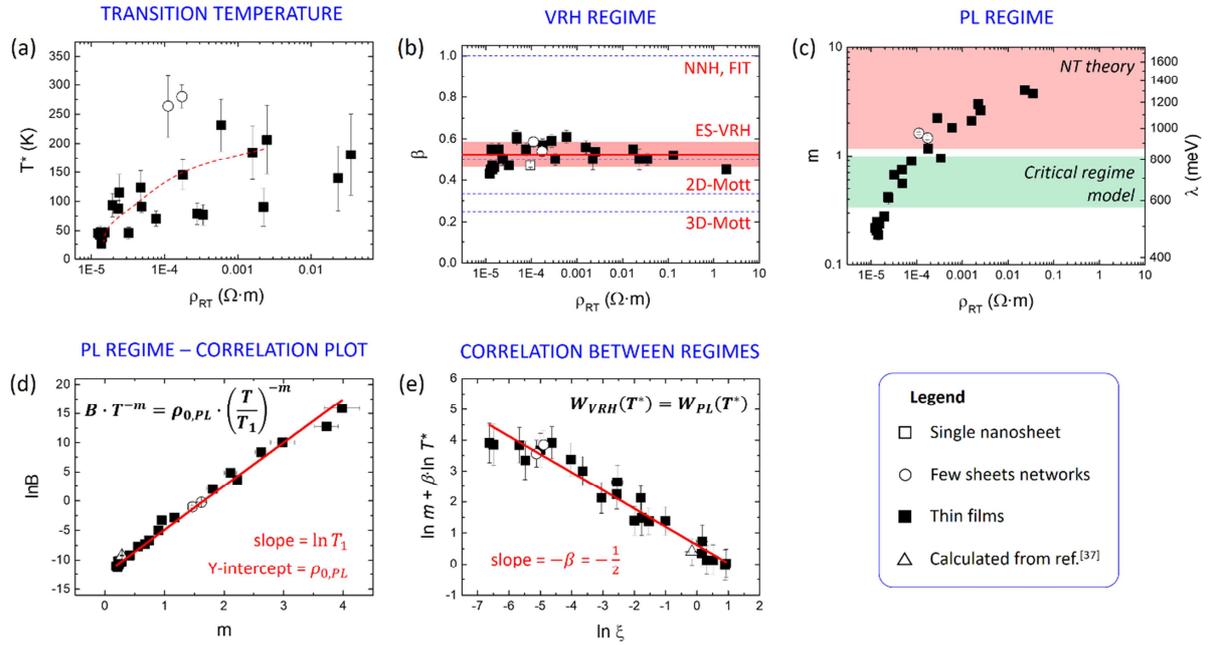

**Figure 3.** Behaviour of the main parameters *vs* room temperature resistivity $\rho_{RT}$. (a) Transition temperature T* *vs* RT resistivity for all samples. The red dashed line is a guide to the eye. (b) Values of exponent $\beta$, as calculated for low temperature data corresponding to the VRH regime. The average $\beta$ value (red line) measured on the 28 devices is 0.52±0.06, corresponding to the ES-VRH model. The shadowed area width corresponds to twice the standard deviation. For the sake of comparison, dashed blue lines show the values expected for other regimes: VRH in 2D and 3D ($\beta$ = 1/3 and $\beta$ = 1/4, respectively) and Arrhenius (NNH) ($\beta$ = 1). (c) Values of power exponent m corresponding to the PL regime, calculated from high temperature data, plotted *vs* RT resistivity. The corresponding relation energy $\xi$ is also reported on the secondary Y axis. The shadowed areas corresponded to the range of values calculated with critical regime model (green) and reported in literature for conjugated polymers using NT theory (red). $\beta$ and *m* parameters shown in in b) and c) are calculated from reduced activation energy. Low signal-to-noise ratio affected devices with $\rho_{RT}$ > 0.1 Ω·m preventing the calculation of T* and *m*

parameters. Correlation plots. (d) PL regime: ln $B$ *vs* m obtained from eqn. 1 for high temperature data. The achieved linear trend (red line) indicates that all the RGO networks follow the same CT mechanism with a given characteristic energy ($k_B T_1$). (e) Continuity of W(T) curve at T*. $\xi$ is directly calculated for ES-VRH using the parameter $T_0$. The linear fit (red line) is calculated using eqn. 3 and the obtained mean $\beta$ value amounts 0.55±0.03 (line slope), in excellent agreement with the arithmetic mean value calculated in figure 3b. In all graphs, symbols correspond to different type of devices: (■) thin films, (○) few sheets partially overlapped, (□) single high-reduced RGO sheet and (△) thin films from ref.[37]. Full details of each device are in SI, table S3 and S4.

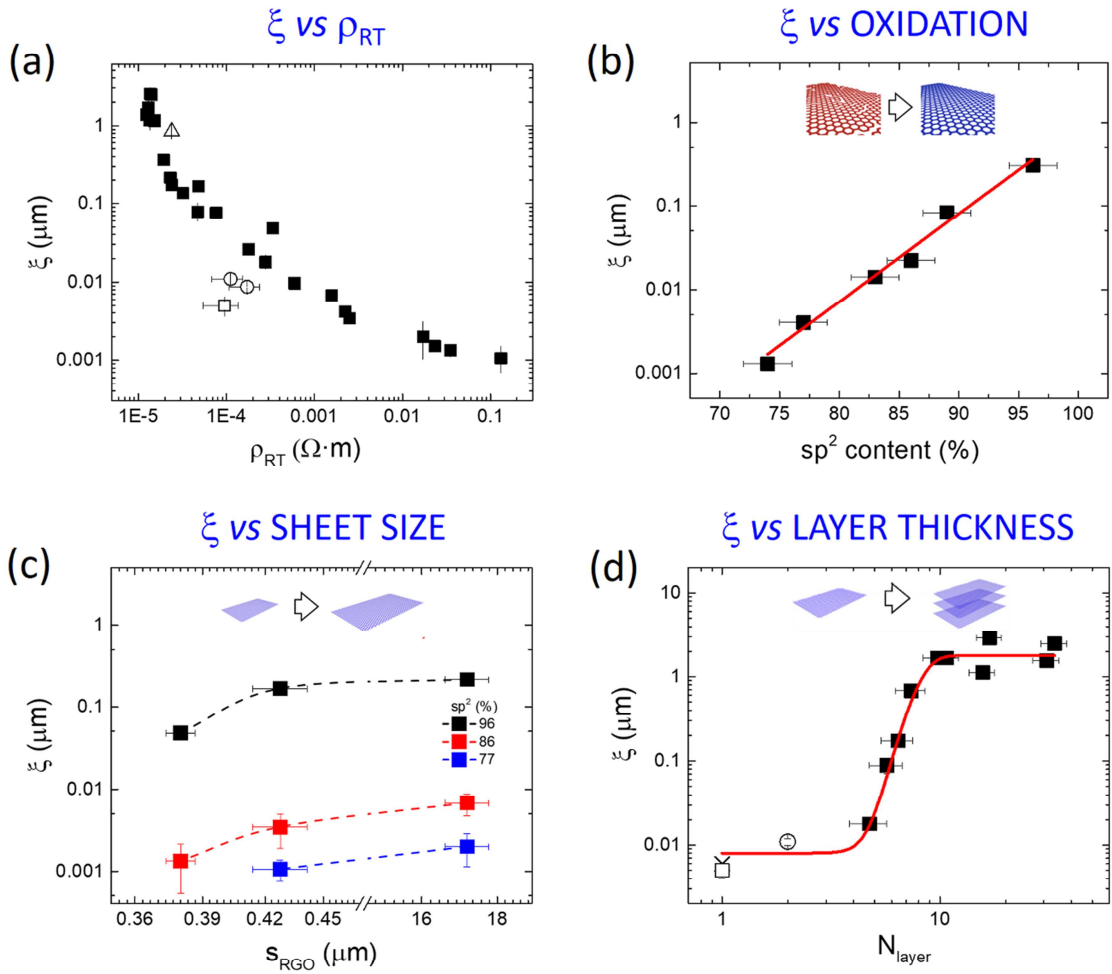

**Figure 4.** Localization length (ξ) as function of: (a) the room temperature resistivity; (b) $sp^2$ content in the nanosheets, tuned by thermal reduction, (c), sheet size ($s_{RGO}$) tuned by sonication of the nanosheets before processing, and (d) film thickness ($N_{layer}$). Data in (b) are fitted with an exponential function (red line in semi-logarithmic plot). In (c) three aromatic $sp^2$ content (■ 77%, ■ 86% and ■ 96%) are analyzed for each sheet lateral size. Lines are just a guide for the eye. In (d) different symbols indicate different devices. In (a) and (d), symbols correspond to different type of devices: (■) thin films, (○) few sheets partially overlapped and (□) single high-reduced RGO sheet. For a detailed description of all devices see table S3. Red line corresponds to the curve calculated with eqn. S6.3b. The size of a typical $sp^2$ region taken from ref.[28] is also reported ($\times$ symbol) for comparison.

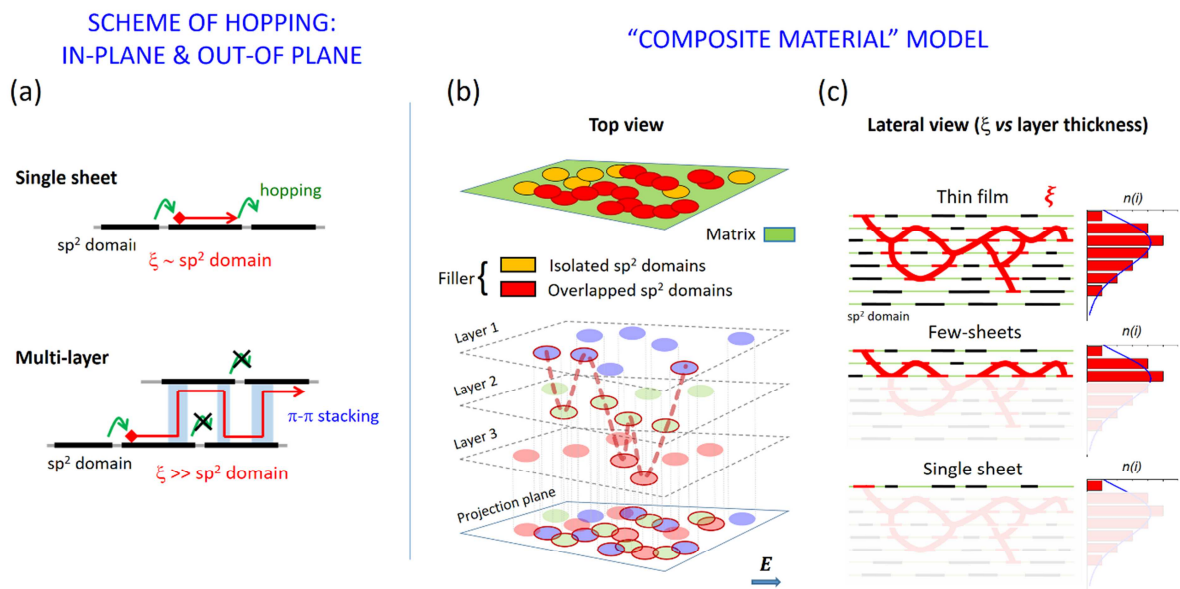

**Figure 5.** Cartoons representing the charge transport in RGO networks. (a) CT in single RGO sheet (pure 2D transport) described by semimetallic behavior achieved in the aromatic domain and hopping at the edges (voids or defects). In the case of stacked RGO sheets, charges can move in 3 dimension due to π-π stacking allowing to circumvent the defect. "Composite material" model: (b) Top view and (c) Lateral view. (Top view) Three-layer RGO thin film. Each plane is represented as a patchwork of isolated $sp^2$ domains (circles) separated by domain border defects composed of voids, C-O functionalities or other defects. For sake of clarity, we distinguish each layer with a different colors. Dashed red line corresponds to a random path connecting overlapped disks. (Lateral view) Dependence of $\xi$ with number of layers. In the case of the thin films $\xi$ extend up to several $sp^2$ domains forming a random path with multiple passes

and dead ends. Reducing the film thickness, the length reduces to the ultimate case of half the size of the sp$^2$ single domain in the case of single layer. The corresponding number of steps *n(i)* (bars) follow the Poisson distribution (blue lines).

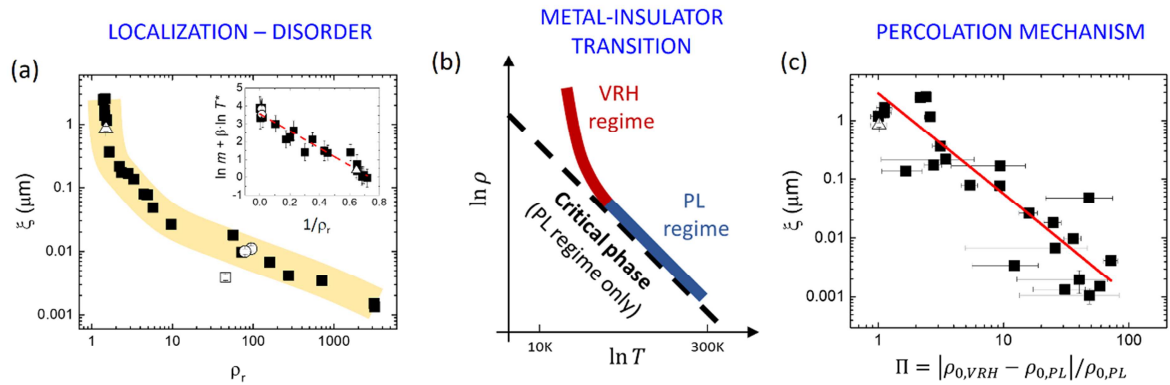

**Figure 6.** CT of RGO networks. (a) The localization length calculated by applying the ES-VRH to experimental data, plotted *vs* the disorder degree ($\rho_r$). The yellow line is a guide for the eye. Inset: correlation plot between the right expression of eqn.4 – as calculated by PL regime – and $1/\rho_r$. (b,c) Percolation-type description of the Metal-Insulator Transition (MIT). (b)Schematic diagram of resistivity *vs* temperature in the vicinity of MIT. A typical measured ρ(T) is reported as a continuous curve (the red and blue sections indicate the different regimes) while a pure critical phase is reported as a black dashed curve. (c) Plot of the localization length ξ *vs* the parameter Π in log-log scale, suggesting the presence of percolation. The slope of the linear fit (red line) is 1.7±0.2.

ASSOCIATED CONTENT

**Supporting Information.** List of devices; Compositional analysis RGO thin films (XPS); images of RGO devices thin films and Single RGO and RGO few-sheets networks; ρ(T) equations, Charge transport properties of RGO thin films reported in literature; Comments on W(T) analysis; ρ(T) and W(T) plot of all devices; correlation plots; experimental parameters of all the prepared devices; continuity of function W at T*. Proof of equation 3.; continuity of function ρ at T*; RGO films as an amorphous polymer network material; disorder and resistivity ratio.

AUTHOR INFORMATION

**Corresponding Author**

* Andrea Liscio: andrea.liscio@artov.imm.cnr.it

* Vincenzo Palermo: vincenzo.palermo@isof.cnr.it

**Author Contributions**

A.L. devised the project, the main conceptual ideas and proof outline. A.K. worked out almost all of the experimental parts, performing the electrical measurements with A.V. and A.B., and the data analysis. A.K. performed XPS measurement. S.D. performed the AFM measurements. Microelectrodes fabrication and electrical measurements are performed by M.G.+P.S. and S.L-A +K.H.K, respectively. A.K., A.C., A.L., M.A. and V.P. have discussed the results and written the article.

**Notes**

The authors declare no competing financial interest.

ACKNOWLEDGMENT

The authors acknowledge the financial support of the Graphene Flagship Core 3 project (GA-881603). This work was partially supported by European Community through the FET-Proactive Project MoQuaS (GA- 610449), the FLAG-ERA JTC 2017 MECHANIC, the ERC Advanced Grant SUPRA2DMAT (GA-833707), by the Italian Ministry for Research (MIUR) through the Futuro In Ricerca (FIR) grant RBFR13YKWX, by Chalmers University of Technology Foundation (Sweden) within the "Rune Bernhardsson's Graphene fund" and the Agence Nationale de la Recherche (France) through the Labex projects CSC (ANR-10-LABX-0026 CSC) and NIE (ANR-11-LABX-0058 NIE) within the Investissement d'Avenir program (ANR-10-120 IDEX-0002-02). Authors are particularly grateful to Emanuele Treossi for providing GO, Rita Tonini for the scientific assistance with use the High Vacuum oven for the heat treatment, Fabiola Liscio for some supporting measurements.

TOC figure

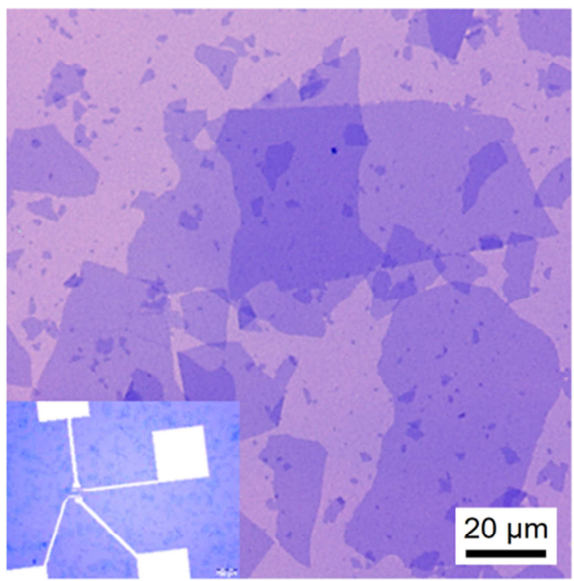
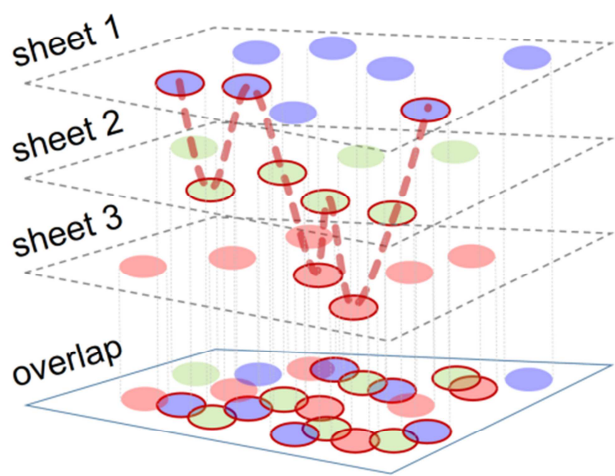

Supporting Information

# Multiscale charge transport in van der Waals thin films: reduced graphene oxide as case study

*Alessandro Kovtun, Andrea Candini, Anna Vianelli, Alex Boschi, Simone Dell'Elce, Marco Gobbi, Kyung Ho Kim, Samuel Lara Avila, Paolo Samorì, Marco Affronte, Andrea Liscio,\* Vincenzo Palermo*[*]

# Contents

# 1. Experimental details

## 1.1 List of devices

- RGO thin films (devices #1 – #25);
- Few-sheets networks (devices #26, #27);
- Single RGO sheet (device #28).

## 1.2 Compositional analysis RGO thin films

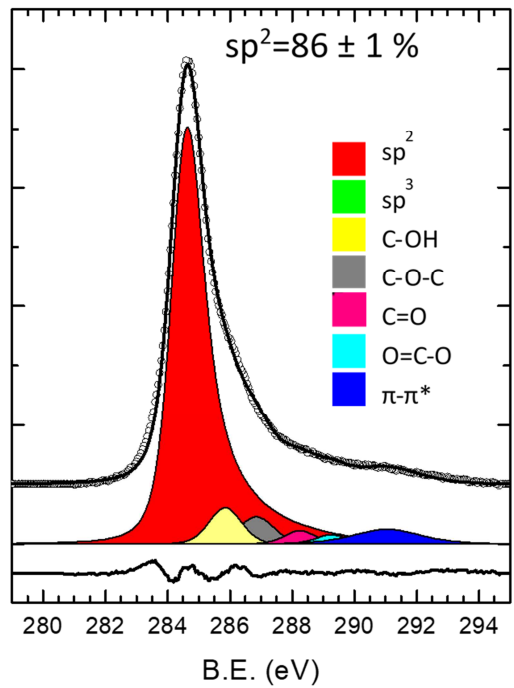

**Figure S1.** Example of the XPS spectrum of the C 1s signal (device #14), thickness = 8±1 RGO layers, $T_{ann}$ = 600 °C. Aromatic $sp^2$ carbon (red area, 86% of the total amount of carbon signal) and residual defects: $sp^3$ hybridization (green), hydroxyl (C-OH, yellow), epoxy (C-O-C, grey), carbonyl (C=O, magenta) and carboxyl (O-C=O, cyan) groups. Spectrum is published in our previous work.[1]

## 1.3 RGO thin films

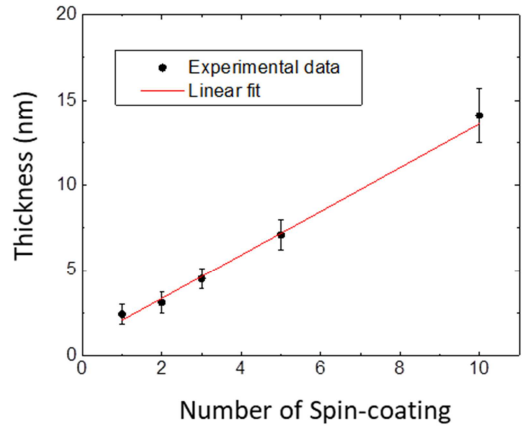

**Figure S2.** Thickness of RGO thin films (after thermal annealing $T_{ann}$ = 900°) *vs* number of spin-coating. The measured values (black dots) increase linearly. The slope of the linear fit (red line) amounts 1.3 nm/spin-coating.

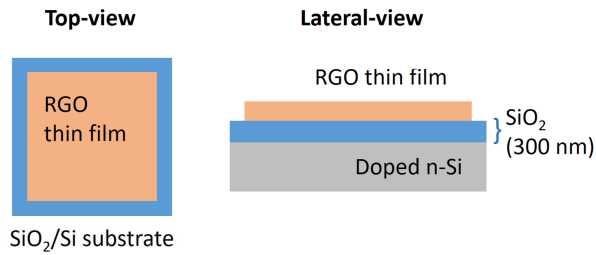

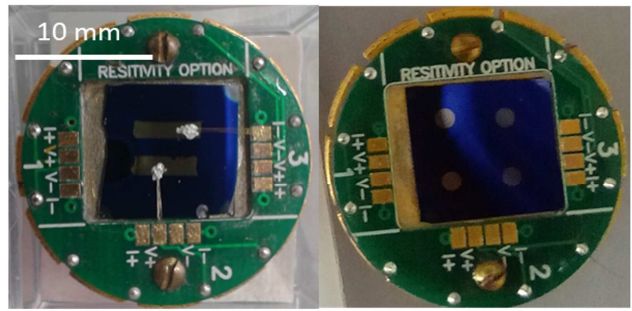

**Figure S3:** (a) Scheme of the samples. (b) Optical microscope pictures of two representative devices in the 2 probes (device #3) and in the 4 probes (device #6) geometry as mounted on the experimental set up. Device geometry: 2 probes: channel length = 1 mm; channel width = 6 mm; 4 probes: 4 circular gold electrodes (1 mm diameter) forming a square (6 mm size).

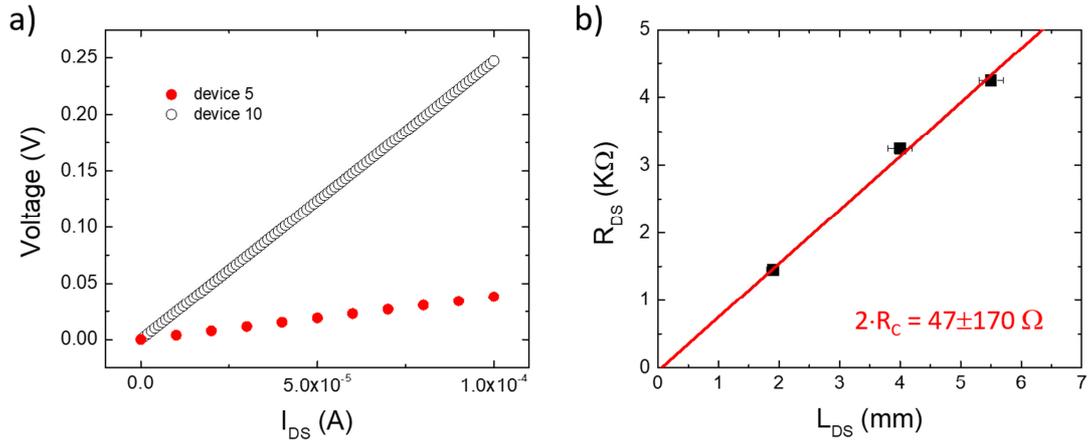

**Figure S4.** (a) Representative linear IV characteristics of the measured devices: (device #10) 4-probes Van der Pauw geometry in which current is biased while measuring the voltage and (device #5) 2-probes in which voltage is biased while measuring the current. In all the devices is observed the linear (Ohmic) regime. (b) Calculation of the contact resistance $R_C$ using Transfer Line Method (TLM). Representative analysis. $R_{DS} = 2 \cdot R_C + \rho_{film} \cdot \frac{w}{h} \cdot L_{DS}$, where the channel width w = 8 mm, the film thickness h = 3.7±0.7 nm. RGO film was annealed at 900°C (sp$^2$ content = 96%). The $R_C$ value calculated from the linear fit amounts 25±80 Ω. Such value can be easy neglected and set to zero because: i) amounting two orders of magnitude lower than the measured resistance (kΩ range) and ii) being lower than the experimental confidence.

### 1.4   Single RGO and RGO few-sheets networks

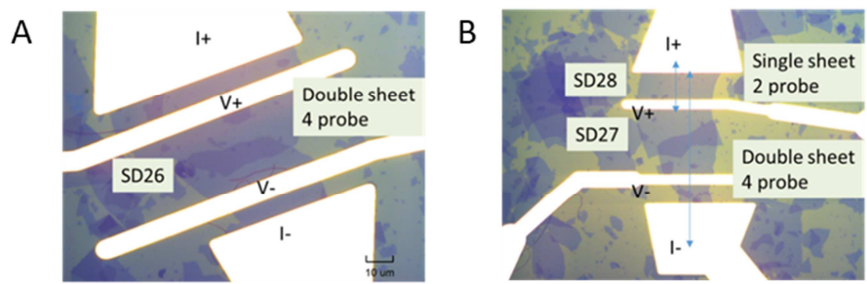

**Figure S5.** Optical Microscope image of devices (A) #26 and (B) #27, #28. Single Sheet configuration was measured with 2 probe configuration, double sheet with 4 point probe.

## 2. Charge transport models

### 2.1 ρ(T) equations, a brief overview

At low temperature, charge transport in disordered materials is typically occurring *via* charge hopping in a disorder-broadened density of states near the Fermi level, g(E$_F$),[2] as demonstrated in previous works (a comprehensive list is reported in the SI). In the Ohmic regime, the resistivity $\rho(T)$ can be modelled by a stretched exponential behavior:

$$\rho(T) = \rho_{0,VRH} \cdot exp\left\{\frac{T_0}{T}\right\}^{\beta} \qquad (S2.1)$$

where $\rho_{0,VRH}$ is a prefactor and $\beta$ a characteristic exponent. $T_0$ represents a characteristic temperature correlated to the localization length ($\xi$), the higher the first one, the lower the latter. The analytic expression depends on the model, as reported in Table S1. $\xi$ is defined as the average spatial extension of the charge carrier wave function: the lower the $T_0$, the larger the $\xi$. The stretching exponent $\beta$ is strongly dependent on the shape of g(E$_F$) and can be used to infer the correct CT model for the system under study, as detailed below. The functional form of equation 1 is quite general, applicable to different models commonly referred as Variable Range Hopping (VRH). When the density of states is constant (Mott-VRH model),[3-4] the $\beta$ value directly depends on the system's dimension (D) with the form $\beta = 1/(D + 1)$. In 1975 A. L. Efros and B. I. Shklovskii suggested the existence, at low-enough temperature, of a gap at the Fermi level due to the Coulomb interaction between the occupied, excited state above E$_F$ and the hole left by the same electron below E$_F$.[5] In the Efros-Shklovskii (ES) model, the hopping of an electron from a state with energy $E_i$ to a state with energy $E_j$ should imply a positive energy change of the systems of $\Delta H(i \to j) = E_j - E_i - e_{ij} > 0$, where the last term describes the excitonic effect, *i.e.* the Coulomb interaction of the electron-hole pair just created. This implies that, at the Fermi level, there should be a gap $>e_{ij}$ between any two states *i* and *j* for the inequality to be true. As a consequence of this, in the ES variable range hopping regime the characteristic exponent $\beta$ of equation 1 is always $\beta=1/2$ and does not depend on the system dimensionality.[6] When CT proceeds through thermally activated hopping between nearest-neighbours (NNH model),[7] it can be modelled with an Arrhenius expression, giving $\beta = 1$.

A further model describing CT is the so-called Fluctuation-Induced Tunneling (FIT) transport.[8] The model was developed to describe the conduction mechanism in disordered systems, *i.e.* conducting polymers and nanocomposite materials, featuring metal pathways separated by small insulating barriers, and recently it has been considered for disordered materials with metallic and semiconducting regions. Also, FIT can be described by equation1 introducing an effective temperature $T_{eff} = T + \Delta$, where $\Delta$ is a characteristic temperature, and $\beta = 1$.

$$\rho(T_{eff}) = \rho_{0,FIT} \cdot exp\left\{\frac{T_0}{T+\Delta}\right\} = \rho_{0,FIT} \cdot exp\left\{\frac{T_0}{T_{eff}}\right\} \tag{S2.2}$$

Below $T_0$ the conduction is dominated by the tunneling of carriers through the barrier, at metal insulating junction, and $\Delta$ is the temperature above which the thermally activated conduction over the barrier begins to occur and are defined as

$$T_0 = \frac{8\varepsilon_0}{e^2 k_B}\left(\frac{AV_0^2}{w}\right); \Delta = \frac{16\varepsilon_0 \hbar}{\pi(2m_e)^{1/2}e^2 k_B}\left(\frac{AV_0^{3/2}}{w}\right) \tag{S2.3}$$

where $\varepsilon_0$ is the vacuum permittivity, e is the electronic charge, $2\pi\hbar$ is Planck's constant, and $m_e$ is the electronic mass. The width and height of the potential barrier are $w$ and $V_0$, with junction area of $A$.

The characteristic parameters of all CT models described above are summarized in Table S1.

**Table 1.** Characteristic parameters, as defined by Mott-, ES- VRH, Arrhenius and FIT models. *a* is a numerical constant, $\varepsilon_0$ and $\varepsilon_r$ represent the vacuum permittivity and the relative permittivity of the material, $g(E_F)$ the DOS at the Fermi energy and $k_B$ the Boltzmann constant. $E_{act}$ is the activation energy between nearest-neighbour sites responsible of the conduction. In the case of FIT model, w and $V_0$ are the width and height of the potential barrier, with junction area $\Sigma$.

|   | Mott-VRH | ES-VRH | NNH (Arrhenius) | FIT |
|---|---|---|---|---|
| $\beta$ | $\frac{1}{D+1}$ | $\frac{1}{2}$ | 1 | 1 |
| $T_0$ | $\frac{a}{g(E_F) \cdot k_B \cdot \xi^D}$ | $\frac{2.8\, e^2}{4\pi\, \epsilon_0 \epsilon_r \cdot k_B \cdot \xi}$ | $\frac{E_{act}}{k_B}$ | $\frac{8\epsilon_0}{e^2 k_B}\left(\frac{\Sigma V_0^2}{w}\right)$ |

A further "robust" function form describing the temperature-dependence of the electrical resistivity is the so-called power law (PL): $\rho = B \cdot T^{-m}$. This is an empirical function, based on experimental data, where *B* is a scale factor that does not depend on the temperature.

This empirical behaviour is quite general, being observed in different systems, *i.e.* highly conductive systems close to the metal-insulator transition, inorganic semiconductors and organic polymers, and in a wide range of temperatures. A phenomenological curve correlating the power-law behaviour with both temperature and applied bias (*i.e.* ohmic and non-ohmic regimes) was developed recently and it is commonly called *universal scaling curve*.[9] Several theoretical frameworks try to describe the PL behaviour and no consensus exists (see introduction of ref.[10] for a detailed list). We thus used a general approach, in analogy with VRH formula shown in equation S2.1 above, describing the PL regime in the case of low bias as:

$$\rho(T) = B \cdot T^{-m} = \rho_{0,PL} \cdot \left(\frac{T}{T_1}\right)^{-m} \quad \text{(S2.4)}$$

where $\rho_{0,PL}$ is a prefactor and $T_1$ is a term that mathematically allows us to define a correct dimensional analysis and that physically corresponds to the characteristic temperature associated with the CT mechanism.

Historically, most of the studies on CT with PL model focused on studying the exponent *m* only, which is the parameter, describing the functional shape of the measured $\rho(T)$. Conversely, only fewer works investigated the characteristic temperature $T_1$, although the latter can provide direct information on the transport mechanism. In the specific case materials of conjugated materials, different models are required to describe different ranges of m. In the case of electronic transport, for $1/3 < m < 1$ the model describes disordered systems with large localization length, and $T_1$ is equal to the Fermi temperature.[11-12] When $m \geq 1$, (integer values) the transport is characterized by a single-/multi-phonon assisted process and *m* corresponds to the number of phonons involved in the process with characteristic energy = $k_B T_1$. For instance, in the case of amorphous carbon films *m* ranges between 15 and 17.[13] In the case of strong electron-lattice coupling, multiphonon processes are described in terms of polaronic transport. In such case, m can range between 0.3 and 11,[14] as observed in disordered semiconducting polymers and described by different models (*i.e.* Nuclear tunnelling, Miller-Abrahams and Marcus theory).

A further aspect related to the wide range of cases described by the equation 2 is that also the stretched exponential form of equation S2.1 can be approximated to a power law in a wide range of parameters. For instance, by numerical simulation Rodin *et al.*[15] showed that equation S2.1 collapsed to equation S2.2 for high doping while Gornyi *et al.*[16] modelled the transition between the two regimes at increasing disorder.

### 2.2 Charge transport properties of RGO thin films reported in literature

The published results lie in a wide range, as summarized in Table S2. This issue can be ascribed to the complexity of the chemical structure of GO, whose properties strongly depend on the mechanisms of reduction, and the lack of a standard procedure to perform quantitative analysis of the R(T) measurements. Moreover, the use of samples with different sizes and geometries, and as well as different temperature ranges, makes the comparison of results complicated.

Regarding the quantitative analysis, different approaches are used to manage of the acquired data. Only few works report a systematic W(T) analysis showing different results (*i.e.* 3D-Mott,[17] 2D-Mott[18] and ES[19] VRH transport). It is noteworthy to underline that no observation of the Mott 2D-VRH is performed[20-21] using the W(T) analysis. Unfortunately, most of the published paper compare a few number of devices in similar conditions – one or two, at least –, only ref.[21-23] have tuned the localization length by changing the reduction degree. Moreover, a fine control of the film thickness is lacking and several works report studies performed on films produced by drop-casting deposition, a method that does not allow to have uniform films effectively.

Summary of charge transport studies performed on RGO thin films.

**Table S2.** Short summary of the published results on the charge transport properties of RGO. The value for the parameter $\beta$ (Eqn. 1) is reported only if it was obtained from the fit of the reduced activation energy. Label "A" in the High T. column indicates that the high temperature data have been modelled with an Arrhenius-like function. Last column indicates the deposition methods: drop-casting (D) or spin-coating (S). RGO films are usually prepared by D method, resulting in a poor control on the final film thickness. On the contrary, S method allows to obtain macroscopic thin films with nanometric control of the thickness, as typically employed in organic electronics and in supra-molecular chemistry.

| ref | Reduction procedure | VRH | Temperature range (K) | $\beta$ | Localization length tuning | Regime at high temperature | Deposition |
|---|---|---|---|---|---|---|---|
| 21 [a] | Hydrazine | 2D | 78-240 | - | By reduction degree | A | D |
| 20 [a] | Hydrazine | 2D | 27-200 | - | No | - | D |
| 22 [a] | Hydrazine | ES | 50-221 | 0.48 | By reduction degree | - | D |
| 24 [a] | T. anneal. | ES | 10-70 | 0.5 | No | - | D |
| 18 | Vitamin C | 2D | 50-180 | 0.32 | No | A | D |
| 19 | Electrochem. | ES | 10-250 | 0.46 | No | - | D |
| 23 | T. anneal | 2D | 2-100 | - | By reduction degree | A | S |
| 25 | Hydrazine | 3D | 10-300 | - | No | A | D |
| 26 | T. anneal | 3D | 20-250 | - | No | A | D |
| 27 | Hydrazine | 3D | 77-300 | 0.22 | No | - | D |

[a] Single RGO sheet.

## 3. R(T) and W(T) analysis

### 3.1 FIT transport. Comments on W(T) analysis

Taking into account the expression reported in Equation S2.2, the corresponding reduced activation energy of FIT model assumes the form:

$$lnW(T_{eff}) = \alpha - lnT_{eff} \qquad (S3.2)$$

The use of $T_{eff}$ allows to use a clear and compact analytic form, however the experimental parameter is the temperature T and in particular an analytical expansion in term of $lnT$ has to be provided. The logarithmic expressions can be written as:

$$lnT_{eff} = ln(T + \Delta) = lnT\left(1 + \frac{\Delta}{T}\right) = lnT + ln\left(1 + \frac{\Delta}{T}\right) \qquad (S3.3)$$

In the limit T>>Δ, we can use the Maclaurin series and equation S3.6 becomes $lnT_{eff} \cong lnT + \frac{\Delta}{T}$.

Thus, the Equation S3.3 assumes the form:

$$lnW(T_{eff}) \cong A - lnT - \frac{\Delta}{e^{lnT}} \qquad (S3.5)$$

It is noteworthy to remind that the curve described by the Equation S3.5 is a concave function and describes a line in the case Δ = 0 (*i.e.* Arrhenius-like case), as reported in figure S6.

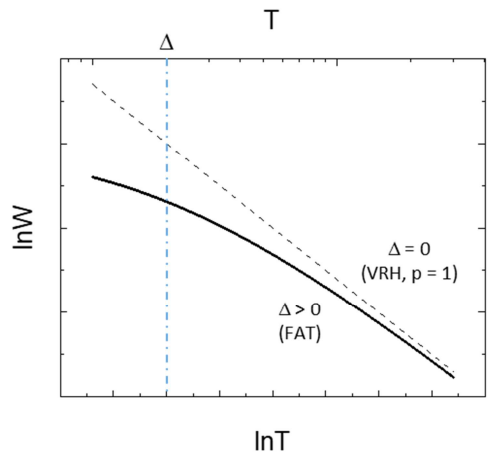

**Figure S6.** Plot of W($T_{eff}$) in lnW *vs* lnT space. FIT regime is described by a concave curve (bold curve) collapsing to a line (dot line) when Δ = 0, corresponding to a VRH-like case with $\beta$ = 1.

## 3.2 ρ(T) and W(T) data

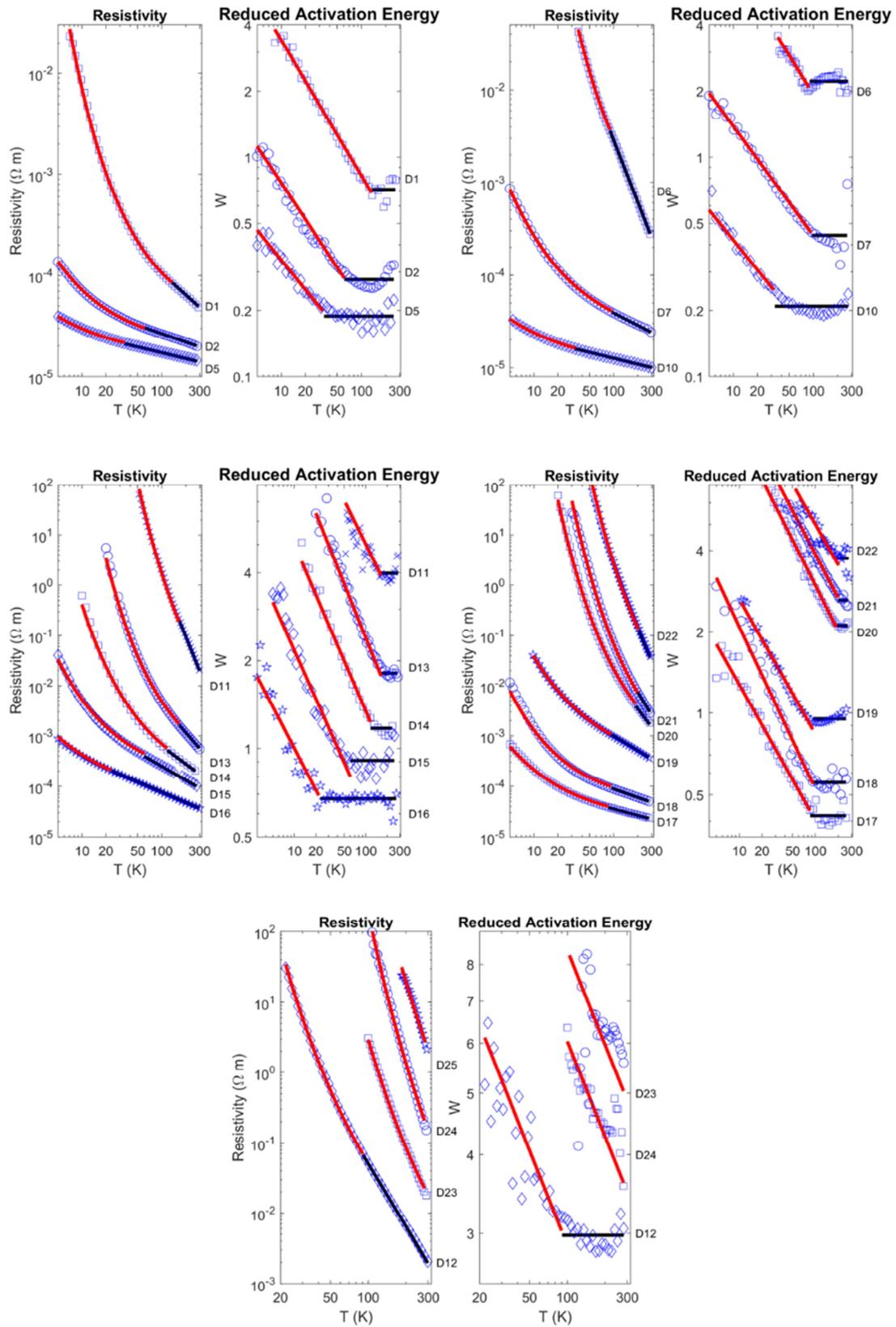

**Figure S7**. Device #1-#26. RGO thin-films.

## 3.3 R(T) and W(T) of RGO few-sheets networks

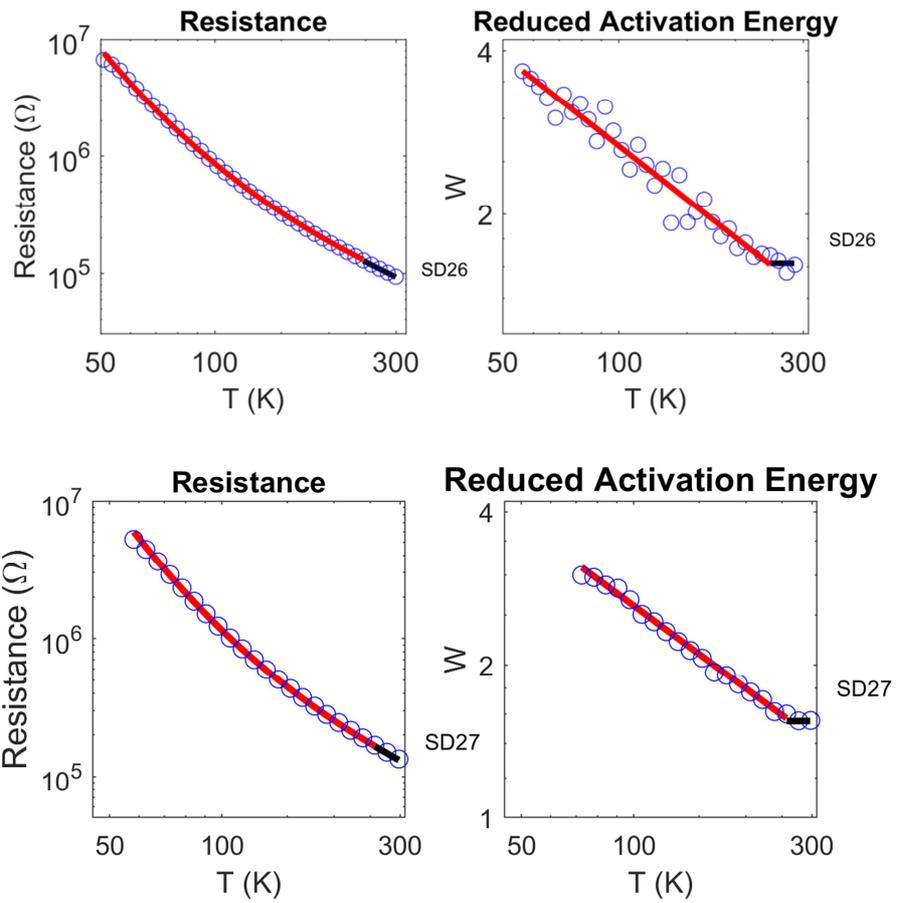

**Figure S8.** Device #26-#27. RGO few-sheets.

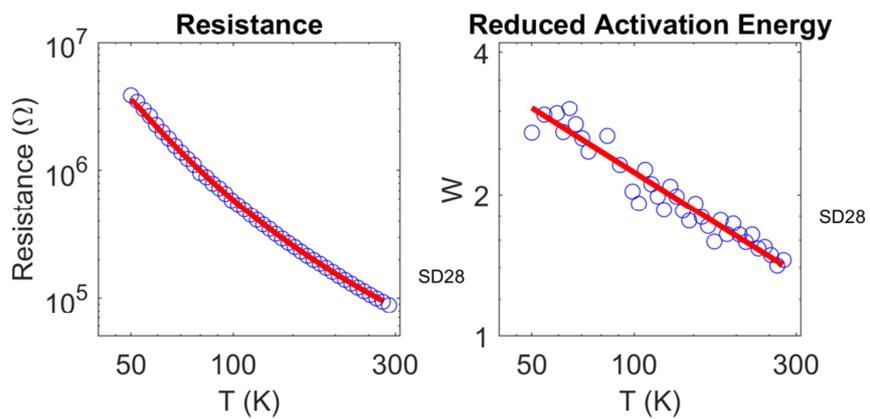

**Figure S9.** Device #28. Single RGO sheet

## 4. Correlation plots

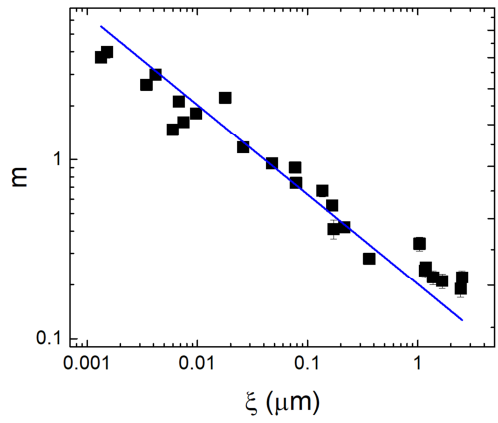

**Figure S10.** *m* parameter *vs* localization length $\xi$. Blue line indicates the best-fit performed using the equation $\xi \propto 1/m^2$ numerically obtained by Rodin for 1D systems.[15]

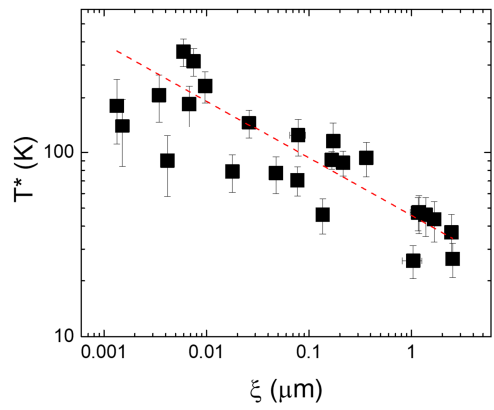

**Figure S11.** Transition temperature (T*) *vs* localization length $\xi$. Dotted red line indicates the linear best-fit in log-log scale.

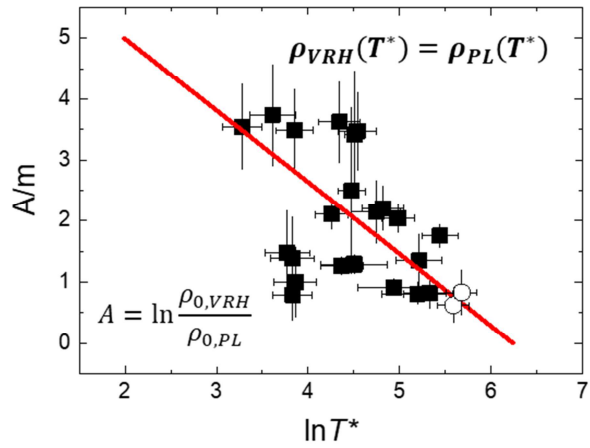

**Figure S12.** Continuity of ρ(T) curve at T* (see paragraph S6.2). The value A reported is the ratio of the prefactors of VRH and PL regimes. The linear fit (red line) is calculated using equation 3 and the slope amounts to -1.1±0.2 (line slope) in agreement with the correlation slope = -1.

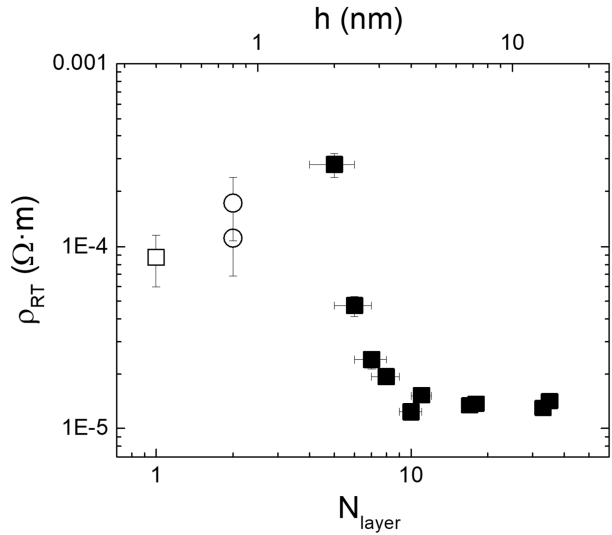

**Figure S13.** Room-temperature resistivity $\rho_{RT}$ (RT = 300 K) *vs* number of layers measured on RGO thin films reduced at $T_{ann}$ = 900°. For clarity is reported the corresponding film thickness (h). $\rho_{RT}$ values measured on devices containing a single of few RGO sheets partially in contact $\rho_{RT}$

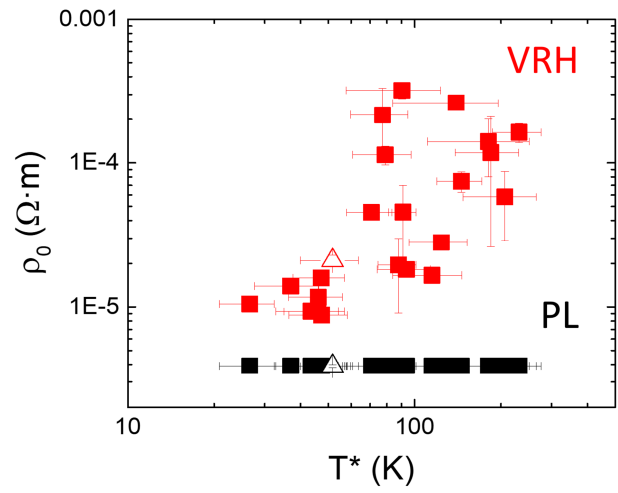

**Figure S14.** Prefactors $\rho_{0,VRH}$ (red) and $\rho_{0,PL}$ (black) *vs* transition temperature ($T^*$).

## 5. Summary of experimental parameters of all the prepared devices

**Table S3.** Comprehensive results of the analysis shown in Figure 1 for all the reported devices. The two regimes are separated at the characteristic temperature $T^*$, as directly obtained by W(T) plots and using equation 3. The carbon $sp^2$ fraction was estimated by XPS analysis, the number of effective layers ($N_{eff}$) by AFM. †The relative error amounts to 5% in log-space (log $T^*$). ‡$W(T)$ of device #25 shows a high-noise level because of the low reduction degree and the corresponding higher resistivity. For this reason, the parameter $\beta$ was fixed to 0.5 in order to evaluate the compatibility of the data with the ES-VRH.

| # | $sp^2$ | $N_{eff}$ | $\beta$ | $\xi$ (nm) | $T^*$ (K) from W(T) plot† | $T^*$ (K) from equation3 | m |
|---|---|---|---|---|---|---|---|
|  | (±2) |  |  |  |  |  |  |
| 1 | 96 | 6±1 | 0.61±0.03 | 90±20 | 140±30 | 123±28 | 0.74±0.05 |
| 2 | 96 | 8±1 | 0.55±0.03 | 680±110 | 60±11 | 105±20 | 0.28±0.02 |
| 3 | 96 | 11±1 | 0.46±0.02 | 1670±210 | 32±5 | 47±10 | 0.24±0.02 |
| 4 | 96 | 18±2 | 0.47±0.02 | 2900±320 | 33±5 | 27±6 | 0.22±0.02 |
| 5 | 96 | 35±2 | 0.45±0.02 | 2470±230 | 33±6 | 37±9 | 0.19±0.02 |
| 6 | 96 | 5±1 | 0.59±0.03 | 18±2 | 90±18 | 79±18 | 2.22±0.05 |
| 7 | 96 | 7±1 | 0.50±0.03 | 170±10 | 100±20 | 115±31 | 0.41±0.05 |
| 8 | 96 | 10±1 | 0.43±0.02 | 1670±210 | 35±5 | 46±11 | 0.22±0.02 |
| 9 | 96 | 17±1 | 0.55±0.02 | 1120±190 | 35±5 | 47±11 | 0.25±0.02 |
| 10 | 96 | 33±2 | 0.45±0.02 | 1550±180 | 35±5 | 44±11 | 0.21±0.02 |
| 11 | 74 | 9±1 | 0.5±0.1 | 1.3±0.1 | 160±35 | 140±56 | 4.0±0.3 |
| 12 | 77 | 9±1 | 0.5±0.1 | 4.0±0.2 | 90±18 | 90±33 | 3.0±0.2 |
| 13 | 83 | 9±1 | 0.61±0.03 | 14±2 | 150±33 | 231±44 | 1.8±0.1 |
| 14 | 86 | 9±1 | 0.57±0.03 | 22±2 | 120±25 | 145±26 | 1.17±0.05 |
| 15 | 89 | 9±1 | 0.55±0.03 | 83±5 | 60±11 | 71±13 | 0.90±0.05 |
| 16 | 96 | 9±1 | 0.47±0.02 | 300±20 | 20±3 | 47±10 | 0.67±0.05 |
| 17 | 96 | 7±1 | 0.50±0.03 | 205±10 | 88±22 | 88±13 | 0.42±0.02 |
| 18 | 96 | 7±1 | 0.60±0.03 | 170±10 | 90±18 | 91±10 | 0.56±0.02 |
| 19 | 96 | 7±1 | 0.50±0.03 | 48±2 | 90±18 | 78±18 | 0.95±0.05 |
| 20 | 86 | 9±1 | 0.56±0.03 | 6.8±0.5 | 190±43 | 184±46 | 2.1±0.1 |
| 21 | 86 | 7±1 | 0.54±0.03 | 3.4±0.3 | 200±46 | 206±59 | 2.6±0.1 |
| 22 | 86 | 9±1 | 0.50±0.03 | 1.3±0.1 | 200±46 | 181±70 | 3.7±0.2 |
| 23 | 77 | 9±1 | 0.50±0.03 | 0.9±0.1 |  |  | - |
| 24 | 77 | 7±1 | 0.50±0.03 | 0.5±0.1 |  |  | - |
| 25 | 77 | 8±1 | 0.50‡ |  |  |  | - |
| 26 | 96 | 2 | 0.56±0.03 | 11±1 | 246±61 | 314±53 | 1.63±0.05 |
| 27 | 96 | 2 | 0.54±0.01 | 10±1 | 280±20 | 354±60 | 1.54±0.02 |
| 28 | 96 | 1 | 0.45±0.02 | 5±1 |  |  | - |

**Table S4.** Electrical parameters of all devices. The lateral sizes were measured by AFM and labeled as (S1) $<s_{RGO}>$ = 17.2±0.6 µm, (S2) $<s_{RGO}>$ = 428±14 nm and (S3) $<s_{RGO}>$ = 380±7 nm. Channel length (L) and width (w) are in mm for thin film devices (#1-25) and in µm for the few sheets devices (#26-28).* Measured by 4 point probe in Van Der Pauw configuration. ** Measured in 4 point probe in linear configuration.

| # | size | $T_{ann}$ (°C) | w (mm) | L (mm) | h (nm) | $\rho_{RT}$ (Ω·m) |
|---|---|---|---|---|---|---|
| 1 | S2 | 900 | 4.1±0.1 | 1.6±0.1 | 2.4±0.3 | 4.72E-05 |
| 2 | S2 | 900 | 4.9±0.1 | 1.6±0.1 | 3.1±0.3 | 1.94E-05 |
| 3 | S2 | 900 | 4.9±0.1 | 1.6±0.1 | 4.5±0.3 | 1.53E-05 |
| 4 | S2 | 900 | 4.9±0.1 | 1.6±0.1 | 7.1±0.4 | 1.37E-05 |
| 5 | S2 | 900 | 4.9±0.1 | 1.6±0.1 | 14.1±0.9 | 1.42E-05 |
| 6 | S2 | 900 | 9.0±0.1* | 9.0±0.1 | 2.0±0.3 | 2.79E-04 |
| 7 | S2 | 900 | 9.0±0.1* | 9.0±0.1 | 2.7±0.3 | 2.40E-05 |
| 8 | S2 | 900 | 9.0±0.1* | 9.0±0.1 | 4.1±0.4 | 1.23E-05 |
| 9 | S2 | 900 | 9.0±0.1* | 9.0±0.1 | 6.6±0.4 | 1.34E-05 |
| 10 | S2 | 900 | 9.0±0.1* | 9.0±0.1 | 13.0±0.9 | 1.30E-05 |
| 11 | S2 | 200 | 3.0±0.1 | 0.85±0.05 | 6.3±0.4 | 2.33E-02 |
| 12 | S2 | 300 | 3.0±0.1 | 0.8±0.1 | 6.3±0.4 | 2.23E-03 |
| 13 | S2 | 500 | 3.0±0.1 | 0.65±0.05 | 5.4±0.4 | 5.94E-04 |
| 14 | S2 | 600 | 3.0±0.1 | 1.3±0.1 | 5.4±0.4 | 1.78E-04 |
| 15 | S2 | 700 | 3.0±0.1 | 0.85±0.05 | 4.5±0.3 | 7.65E-05 |
| 16 | S2 | 940 | 3.0±0.1 | 1.2±0.1 | 3.2±0.3 | 3.20E-05 |
| 17 | S1 | 900 | 5.0±0.1 | 2.0±0.1 | 2,8±1.5 | 2.31E-05 |
| 18 | S2 | 900 | 5.0±0.1 | 2.0±0.1 | 2.8±1.5 | 4.83E-05 |
| 19 | S3 | 900 | 5.0±0.1 | 2.0±0.1 | 2.8±1.5 | 3.36E-04 |
| 20 | S1 | 600 | 5.0±0.1 | 2.0±0.1 | 4.5±1.5 | 1.58E-03 |
| 21 | S2 | 600 | 5.0±0.1 | 2.0±0.1 | 4±2 | 2.50E-03 |
| 22 | S3 | 600 | 5.0±0.1 | 2.0±0.1 | 3.5±1.5 | 3.50E-02 |
| 23 | S1 | 300 | 5.0±0.1 | 2.0±0.1 | 5.4±3.5 | 1.71E-02 |
| 24 | S2 | 300 | 5.0±0.1 | 2.0±0.1 | 4.2±1.5 | 1.31E-01 |
| 25 | S3 | 300 | 5.0±0.1 | 2.0±0.1 | 4.8±2.5 | 1.88E+00 |
| 26 | S2 | 900 | 42±8 µm | 27±5 µm | Two-sheets partially overlapped | 1.11E-04 |
| 27 | S2 | 900 | 40±8 µm | 25±5 µm | Two-sheets partially overlapped | 1.73E-04 |
| 28 | S2 | 900 | 28±5 µm | 10±2 µm | single sheet | 9.52E-05 |

## 6. Mathematical appendix

### 6.1 Continuity of function W at T*. Proof of equation 3.

It is a phenomenological evidence that there is abrupt variation in ρ(T) and W(T) at the transition temperature. For this reason, we assume the equivalence:

$$W_{VRH}(T^{*-}) = W_{PL}(T^{*+}) \tag{S6.1a}$$

The equivalence is conserved using the logarithms:

$$\ln W_{VRH}(T^{*-}) = \ln W_{PL}(T^{*+}) \tag{S6.1b}$$

Thus, by taking into account the analytic form of W of the two regimes we obtain:

$$-\beta \ln T^* + \ln \beta + \beta \ln T_0 = \ln m \tag{S6.1c}$$

Using the formula $\gamma = T_0 \cdot \xi$, as reported in Table 1, we obtain the form shown in equation 3.

$$\beta \cdot \ln T^* + \ln m = -\beta \cdot \ln \xi + \beta \cdot \ln \gamma + \ln \beta \tag{S6.1d}$$

### 6.2 Continuity of function ρ at T*. Proof.

As performed in Section 6.1, we assume the equivalence of the resistivity functions:

$$\rho_{VRH}(T^{*-}) = \rho_{PL}(T^{*+}) \Rightarrow \rho_{0,VRH} \cdot \exp\left\{\frac{T_0}{T^*}\right\}^\beta = \rho_{0,PL} \cdot \left(\frac{T^*}{T_1}\right)^{-m} \tag{S6.2a}$$

and as well as their logarithms. Thus, the eq. S6.2a can be written as:

$$\left\{\frac{T_0}{T^*}\right\}^\beta = -m \cdot \ln T^* + m \cdot \ln T_1 - A, \text{ where } A = \ln\frac{\rho_{0,VRH}}{\rho_{0,PL}} \tag{S6.2b}$$

Using the formula $\gamma = T_0 \cdot \xi$ and calculating the logarithms of the two terms we obtain:

$$-\beta \cdot \ln T^* - \beta \cdot \ln \xi + \beta \cdot \ln \gamma = \ln(-m \cdot \ln T^* + m \cdot \ln T_1 - A) \tag{S6.2c}$$

Using the eq. S6.1d, the first term of eq. S6.2c assumes the form: $\ln m - \ln \beta$. Thus, we calculate the exponential forms obtaining the equivalence:

$$m \cdot \left(\ln\frac{T_1}{T^*} - \frac{1}{\beta}\right) = A \tag{S6.2d}$$

In the case $m = 0$ we obtain a trivial solution as: $A = 0$, corresponding to $\rho_{0,VRH} = \rho_{0,PL}$

### 6.3 RGO films as an amorphous polymer network material

Using the approach to describe the RGO thin films in term of composite materials, the localization length is defined as the product of the number of distinct overlapped sp2 regions ($n_{tot}$), called nodes, visited by a single random path and their average distance ($d$): $\xi \sim d \cdot (n_{tot} - 1)$, where $d$ is the interlayer distance. The nodes (n) are scattered in different layers following the Poisson distribution: $Pois(n) = \frac{n^i \cdot e^{-n}}{i!}$, where $i$ is the index of layer. Thus, the path length can be written as:

$$\xi \sim d \cdot (n_{tot} - 1) = \frac{\phi}{2} + d \cdot \sum_{i=2}^{N_{layer}} n(i) = \frac{\phi}{2} + d \cdot \left[ \frac{\Gamma(N_{layer}+1, n)}{N_{layer}!} - n \cdot e^{-n} \right] \tag{S6.3a}$$

where the cumulative function of the Poisson distribution is the incomplete gamma function $\Gamma(x, y)$.[28] In the case of 3D system (thick RGO film) the localization length amounts in the order of magnitude of microns. Taking into account that the interlayer distance is lower than 1 nm, the number of nodes is ca 1,000. In the limit n ≈ 1,000, the Poisson distribution can be approximated to the Gaussian one. In this case, taking into account that where the comulative distribution of the Gaussian one is $CDF = \frac{1}{2} + \frac{1}{2} \cdot Erf(x)$, the equation S6.3a assumes the form:

$$\xi \sim \frac{\phi}{2} + d \cdot Erf(i - i_0) \tag{S6.3b}$$

where $i_0$ is a free parameter calculated by fit corresponding to the inflection point of the curve, $\xi_{\parallel,0}$ is the localization length related to the single sp$^2$ cluster corresponding to the case of single layer.

### 6.4 Disorder and resistivity ratio

The equation 2 in the main text correlates physical parameters calculated with PL (*m*) and VRH regimes ($\xi$), and T* as well, as demonstrated in the Section S6.1. For clarity, we report equation 3:

$$\beta \cdot \ln T^* + \ln m = -\beta \cdot \ln \xi + \beta \cdot \ln \gamma + \ln \beta \tag{S6.4a}$$

The equation can be written as

$$\beta \cdot \ln T^* + \ln m = -\beta \cdot \ln \xi + \varphi \tag{S6.4b}$$

where $\varphi$ is a constant value. The correlation plot reported in the inset fig 2e clearly shows that:

$$\beta \cdot \ln T^* + \ln m = -1/\rho_r + \varphi' \tag{S6.4c}$$

$\rho_r$ is the resistivity ratio and $\varphi'$ a constant value. By comparing eqns. S6.4b and S6.4c we obtain that

$$\ln \xi \propto 1/\rho_r \tag{S6.4d}$$

Taking into account that $\xi$ depends exponentially on disorder, the equation S6.4d assumes the form:

$$1/\rho_r \propto \quad \text{disorder} \tag{S6.4e}$$